# Influence of chemical substitution and sintering temperature on the structural, magnetic and magnetocaloric properties of $La_{1-x}Sr_xMn_{1-y}Fe_yO_3$

N. Brahiti [1], M. Balli [2], M. Abbasi Eskandari [1], A. El Boukili[3,4], P. Fournier [1]

[1]Institut quantique, Regroupement québécois sur les matériaux de pointe et Département de physique, Université de Sherbrooke, Sherbrooke, J1K 2R1, Québec, Canada

[2] AMEEC Team, LERMA, College of Engineering & Architecture, International University of Rabat, Parc Technopolis, Rocade de Rabat-Salé, 11100, Morocco.

[3] LaMCScI Laboratory, B.P. 1014, Faculty of science. Mohammad V University in Rabat, Morocco.

[4] Materials and Nanomaterials Center, MAScIR Foundation, B.P. 10100 Rabat, Morocco.

**ABSTRACT**

The effects of sintering temperature ($T_s$) and chemical substitution on the structural and magnetic properties of manganite compounds $La_{1-x}Sr_xMn_{1-y}Fe_yO_3$ ($0.025 \leq x \leq 0.7$; $y = 0.01, 0.15$) are explored in a search to optimize their magnetocaloric properties around room temperature. A ferromagnetic (FM) to paramagnetic (PM) phase transition is observed at a Curie temperature $T_c$ that can be controlled to approach room temperature by Sr and Fe substitution, but also by adjusting the sintering temperature $T_s$. Accordingly, the magnetic entropy change ($-\Delta S_M$) quantifying the magnetocaloric effect (MCE) presents a peak at or close to $T_c$ that shifts and broadens with both Sr and Fe doping and is further tuned with sintering temperature. Altogether, we show that it is possible to adjust the strength and dominance of the ferromagnetic coupling in these ceramics, but also using disorder as a tool to broaden and adjust the temperature range with significant magnetic entropy change.

**Keywords:** Magnetocaloric effect, manganite perovskite oxides, chemical substitution.

**INTRODUCTION**

The magnetocaloric effect (MCE) has been used for many years to reach very low temperatures [1-5]. Nearly a century ago, changes in nickel temperature when varying the external magnetic field were originally discovered by Pierre Weiss and Auguste Piccard in 1917 during their study of magnetization as a function of temperature and magnetic field near the magnetic phase transition [1, 6]. The observed temperature increase was then called by Weiss and Piccard "*le phénomène magnétocalorique*" (the magnetocaloric phenomenon) [1, 6]. In the late 1920s, Debye in 1926 [7] and Giauque in 1927 [8] independently proposed an additional thermodynamic explanation of the magnetocaloric effect and suggested a refrigeration process to reach low temperatures using adiabatic demagnetization of paramagnetic salts. The concept was experimentally implemented in 1933 by Giauque and MacDougall [9] allowing them to reach 0.25 K using $Gd_2(SO_4)_8 \cdot H_2O$ salts from the temperatures of liquid helium.

The MCE is an intrinsic property of magnetic materials. It relies on a coupling between the spin system and the lattice as a mean to transfer magnetic entropy to or from the lattice, inducing warming or cooling while magnetizing or demagnetizing it. When a magnetic field is applied adiabatically to a ferromagnetic material, the magnetic entropy decreases due to ordering of the spins. This reduction in magnetic entropy is compensated by an increase in the lattice entropy to preserve total entropy [1-5]. As a result, the magnetic material warms up. Reversely, under an adiabatic decrease of the magnetic field, the moments tend to randomize again leading to an increase of magnetic entropy decreasing accordingly the material temperature.

In recent years, cooling applications based on magnetocaloric materials as refrigerants have attracted more attention because of its potential high energy efficiency in contrast to the fluid compression – expansion conventional systems [1-5]. Magnetic refrigeration near room temperature was implemented for the first time in 1976 by Brown who unveiled an innovative and energy-efficient magnetocaloric device working with gadolinium metal as a magnetic refrigerant [10]. It took advantage of a large variation of the magnetic entropy close to the magnetic transition temperature of Gd under an external applied magnetic field change. The MCE in terms of magnetic isothermal entropy change ($\Delta S_M$) can be evaluated from magnetic measurements using the Maxwell relation [1, 11]:

$$-\Delta S_M(T, 0 \to H) = \mu_0 \int_0^H \left(\frac{\partial M}{\partial T}\right)_{H'} dH' \quad (1)$$

Using magnetic isotherms, magnetization as a function of applied magnetic field for successive temperatures, $\Delta S_M$ is found to be maximum for temperatures where $\frac{\partial M}{\partial T}$ is maximum. This occurs generally in the vicinity of the magnetic phase transition: broadening this transition (with disorder) while preserving a large value of $\Delta S_M$ is the target of the present work.

A giant MCE was observed in $Gd_5Si_2Ge_2$ based compounds near room temperature by Pecharsky and Gschneidner [12]. Since then, a large variety of advanced magnetocaloric materials was proposed and explored for room temperature tasks [1, 11-19]. Since the 1990s, the perovskite manganese oxides also called manganites of general formula $R_{1-x}A_x MnO_3$ (R= trivalent rare earth, A= divalent ion) have been a subject of intensive investigations due to their various functional properties such as colossal and giant magnetoresistance, giant piezoelectric properties, and MCE near room temperature [20-24]. With growing A for R substitution, $x$, the same amount $x$ of $Mn^{3+}$ with the electronic configuration $\left(3d, t_{2g\uparrow}^3 e_{g\uparrow}^1, S = 2\right)$ is replaced by $Mn^{4+}$ with the electronic configuration $\left(3d, t_{2g\uparrow}^3 e_{g\uparrow}^0, S = \frac{3}{2}\right)$ [25]. Large carrier mobility and ferromagnetism are promoted from a strong electron transfer between the filled and empty $e_g$ states of nearby $Mn^{3+}$ and $Mn^{4+}$ ions mediated by oxygen 2p states via the double exchange (DE) mechanism [26]. Moreover, the perovskites structure usually show lattice distortions from the ideal cubic structure to orthorhombic and rhombohedral structures that are mainly caused by Jahn-Teller (JT) distortions and the mismatch of the Mn-O and R-O bond lengths [27]. These lattice distortions play a significant role in determining the physical properties of manganites and have been widely studied in this family (see for example Refs. [27, 28] and references therein). Chemical substitution of the rare earth (R) and metal (Mn) sites offers an obvious path to tune the magnetic, transport and magnetocaloric properties of these manganites in an effort to optimize their cooling capacity. For example, a large MCE from polycrystalline $La_{1-x}A_x MnO_3 (A = Ca, Sr, Ba)$ for $x$ = 0.2 and 0.25 was reported by Guo *et al.* [29, 30]. Maximum magnetic entropy changes of about 5.5 J/kg K at 230 K and 4.7 J/kg K at 260 K were obtained under an applied magnetic field change of 1.5 T, respectively.

The magnetic and magnetocaloric properties of nano-sized $La_{0.8}Ca_{0.2}Mn_{1-x}Fe_xO_3$ ($x$ = 0, 0.01, 0.15 and 0.2) manganites prepared by sol-gel method was studied by Fatnassi *et al.* [31]. They reported that the ferromagnetic-paramagnetic transition occurring in these materials is sensitive to iron doping. In addition, a large MCE near $T_c$ is observed. $-\Delta S_M$ under a magnetic field change of 5 T reaches 4.42, 4.32 and 0.54 J/kg K , for $x$ = 0, 0.01 and 0.15, respectively. In a similar context, Barik *et al.* [32] investigated the effect of Fe substitution on the magnetocaloric effect in $La_{0.7}Sr_{0.3}Mn_{1-x}Fe_xO_3$ ($0.05 \leq x \leq 0.2$). It was shown that the Fe substitution gradually decreases both the Curie temperature and the saturation magnetization. They also showed that a $La_{0.7}Sr_{0.3}Mn_{0.93}Fe_{0.07}O_3$ sample exhibits a large magnetic entropy change $\Delta S_M$ that reaches 4 J/kg K under $\Delta H$ = 5 T. This sample exhibits a refrigerant capacity of 225 J/kg and an operating temperature range over 60 K wide around room temperature. In fact, Leung *et al.* [33] were among the first to study the effect of iron substitution in manganites in the mid-70's. They studied the magnetic properties of $La_{1-x}Pb_xMn_{1-y}Fe_yO_3$ compounds, where a ferromagnetic $Mn^{3+} - O - Mn^{4+}$ double-exchange (DE) interaction competes with antiferromagnetic $Fe^{3+} - O - Mn^{3+}$ and $Fe^{3+} - O - Fe^{3+}$ interactions. More recently, Ait Bouzid *et al.* [34], investigated the magnetocaloric effect in $La_{1-x}Na_xMn_{1-y}Fe_yO_3$ compounds. It was shown that the addition of 10% of iron in $La_{1-x}Na_xMn_{1-y}Fe_yO_3$ decreases the Curie temperature and the magnetic entropy change, while the relative cooling efficiency increases. Altogether, these selected studies demonstrate that Fe for Mn substitution can be used to finely control the Curie temperature and the magnitude of the entropy change.

For the present study, we synthesize co-doped manganites $La_{1-x}Sr_xMn_{1-y}Fe_yO_3$ ceramics with extended doping levels up to *x* = 0.7 and study the influence of strontium and iron substitution at the La and the Mn sites simultaneously. We correlate the impacts of these parallel substitutions on the crystal structure, the magnetic properties and the magnetocaloric effect. As we aim to optimize their magnetocaloric properties for eventual applications in proximity to room temperature, the impact of their growth conditions with a focus on the sintering temperature is also explored for each composition.

**EXPERIMENTAL**

Polycrystalline samples of $La_{1-x}Sr_xMn_{1-y}Fe_yO_3$ ($0.025 \leq x \leq 0.7, y = 0.01, 0.15$) were prepared by the conventional solid-state reaction. High-purity oxides or carbonates $La_2O_3$, $Fe_2O_3$, $MnO_2$ and $SrCO_3$ were used as starting materials. Prior to weighing in the appropriate proportions, $La_2O_3$ was preheated overnight at 900°C. These starting materials were then weighted and thoroughly mixed in an agate mortar until homogeneous powders were obtained. All the powders were heated to 1070°C and then to 1120°C in air for 24h with intermediate grinding steps. The powders were pressed into pellets and subjected to heating cycles at 1170°C, 1220°C and 1250°C. The ceramic samples heated in air were slowly cooled to room temperature at the rate of 5°C/min. Structural properties were analyzed from powder X-ray diffraction (XRD) measurements on both the powders and the pellets at every heating steps using a Bruker-AXS D8-Discover diffractometer in the $\theta - 2\theta$ configuration with a CuKα1 source ($\lambda$ = 1.5406Å) over the $2\theta$ range of 10° to 80°. The structural parameters were obtained by fitting the experimental XRD data using the Rietveld structural refinement FULLPROF software applying the Thompson-Cox-Hastings pseudo-Voigt function with axial divergence asymmetry peak shape function and a linear interpolation for background description. The refinements were performed until reaching the convergence as shown by the goodness of fit ($\chi^2$). The surface morphology of the samples was checked by scanning electron microscopy (SEM).

The DC magnetization measurements were performed using a Superconducting Quantum Interference Device (SQUID) magnetometer from Quantum Design. The temperature dependence of the magnetization was measured from 5 to 380 K with a magnetic field of 0.2 T. The MCE evaluated using the magnetic entropy change was estimated from magnetic isotherms measured as a function of temperature (50-380 K) in 0 to 7 T magnetic fields. The specific heat measurements of $x = 0.15$, $y = 0.01$ and $x = 0.35$, $y = 0.01$ samples were carried out from 3 to 375 K at 0 and 7 T and were performed using a Physical Properties Measurement System (PPMS) from Quantum Design.

**RESULTS AND DISCUSSION**

*Structural properties*

X-ray diffraction (XRD) patterns at room temperature of $La_{1-x}Sr_xMn_{1-y}Fe_yO_3$ ceramics pelletized at 1170°C are presented in Figure 1 for various values of $x$, for $y = 0.01$ in (a) and for $y = 0.15$ in (b). It reveals the presence of the manganite phases together with impurity phases that are virtually absent in the samples with a large Fe doping (y = 0.15) except for $x = 0.7$. The spectra reveal the presence of the rhombohedral crystal structure with $R\bar{3}c$ space group for all the samples which is in accordance with the JCPDS card (no. 53-0058) [35]. However, as shown in the XRD pattern of $La_{1-x}Sr_xMn_{0.99}Fe_{0.01}O_3$ ($x <$ 0.35) with a small amount of iron in Fig. 1(a), a splitting of the diffraction peaks at angles at $\sim 40°$, $\sim 52°$, $\sim 58°$ and $\sim 68°$ is an indication that the structure is not purely rhombohedral and includes the orthorhombic ($Pnma$) phase [36-38]. Moreover, when $x \geq$ 0.5 , a mixture of the rhombohedral and tetragonal ($I4/mcm$) phases can be observed. These observations confirm the trend to phase segregation in manganites for large Sr doping [39-41]. It is interesting to observe that all the XRD patterns of $La_{1-x}Sr_xMn_{0.85}Fe_{0.15}O_3$ ($x < 0.7$) with a large iron content show a single rhombohedral phase with no trace of other symmetry (no doublets) and no impurity phase, suggesting that iron may favor a better Sr homogeneity.

At low Sr and Fe doping, additional peaks with small intensities can be attributed to impurity phases, in particular to $Mn_3O_4$. This impurity phase is known to be widely present in manganites compounds with cation vacancies [42]. $Mn_3O_4$ crystallizes in the tetragonal ($I\,41/amd$) phase [42,43] and is expected to contribute as the dominant impurity phase to the magnetic properties at low temperatures as its paramagnetic to ferrimagnetic transition occurs in the range of 40 to 50 K [43,44].

A magnified view of the peak with the highest intensity ($2\theta \approx 32°$) of the same samples is shown in Figure 2 (a) and (b) for $La_{1-x}Sr_xMn_{0.99}Fe_{0.01}O_3$ and $La_{1-x}Sr_xMn_{0.85}Fe_{0.15}O_3$, respectively. The diffraction peak first shifts down in angle when $x$ increases from 0.025 to 0.15 before shifting to higher angle when the Sr concentration is further increased ($x > 0.15$) for both iron contents. This indicates that the lattice parameters increase first with $x$, but then decrease for $x > 0.15$. Substituting $La^{3+}$ ($r_{La^{3+}} = 1.36$ Å) with a larger $Sr^{2+}$ ion ($r_{Sr^{2+}} = 1.44$ Å) [45] should increase the lattice parameters overall and lead to a decrease of peak angle [46, 47]. However, the density of $Mn^{4+}$ is also increasing with $x$. Since the ionic radius of $Mn^{4+}$ ($r_{Mn^{4+}} = 0.53$ Å) is smaller than that of $Mn^{3+}$ ($r_{Mn^{3+}} = 0.645$ Å) [45], the reverse trend of the lattice parameters is also expected as observed previously [48]. In order to fully capture and understand the structural evolution observed in Fig. 2, we turn to a full analysis of our diffraction spectra using Rietveld refinement.

Figure 3 shows an example of Rietveld refinement fits performed for $La_{0.85}Sr_{0.15}Mn_{0.99}Fe_{0.01}O_3$ and $La_{0.85}Sr_{0.15}Mn_{0.85}Fe_{0.15}O_3$. The fits for the other samples are presented in Figure S1 of the supplementary materials. The spectrum for $La_{0.85}Sr_{0.15}Mn_{0.85}Fe_{0.15}O_3$ in Fig. 3(b) is fitted by considering a single rhombohedral phase ($R\bar{3}c$). However, for $\text{La}_{0.85}\text{Sr}_{0.15}\text{Mn}_{0.99}\text{Fe}_{0.01}\text{O}_3$ in Fig. 3(a), the best fit to the spectra is achieved when a mixture of the rhombohedral ($R\bar{3}c$) and the orthorhombic ($Pnma$) phases is assumed together with the $\text{Mn}_3\text{O}_4$ ($I\,41/amd$) impurity phase. This approach is used to determine the fraction of each phase in $\text{La}_{0.85}\text{Sr}_{0.15}\text{Mn}_{0.99}\text{Fe}_{0.01}\text{O}_3$. A similar procedure is used to analyze all the spectra presented in the supplementary materials which allows us to estimate the fraction of the phases as a function of doping.

Figure 4 presents the phase fractions as a function of the nominal Sr doping level for low iron content (y = 0.01) estimated from the Rietveld refinements. We clearly observe a dominant rhombohedral phase for all the samples with a tendency for an increase in the fraction of the high symmetry phases with increasing $Sr^{2+}$ doping level. The reduction in the density of Jahn-Teller $Mn^{3+}$ ions with increasing Sr doping is at the origin of this gradual evolution towards higher symmetry and the disappearance of the orthorhombic phase. Furthermore, the single rhombohedral symmetry observed for the samples with high Fe content (y = 0.15) is another signature of the decreasing influence of lattice distortions when Jahn-Teller $Mn^{3+}$ is substituted by non-Jahn-Teller $Fe^{3+}$. This effect dominates even for the lowest Sr doping (x = 0.025) where even a small amount of $Fe^{3+}$ (y = 0.15) is enough to overcome the impact of the Jahn-Teller distortions driven by the $Mn^{3+}$ cations.

The results of the calculated lattice parameters and unit cell volume ($V$) of the dominant rhombohedral phase by Rietveld refinement for these $\text{La}_{1-x}\text{Sr}_x\text{Mn}_{1-y}\text{Fe}_y\text{O}_3$ ($0.025 \leq x \leq 0.7, y = 0.01, 0.15$) compounds are presented in Table 1 revealing their trends as a function of the Sr and Fe substitution levels. With the definition of B, B' as Mn or Fe, and A as La or Sr with the general formula $ABO_3$, Table 1 includes also the average $\text{La(Sr)} - \text{O}$ distance ($d_{A\text{-O}}$), the average $\text{Mn(Fe)} - \text{O}$ bond length ($d_{B-O}$), the average Mn(Fe) $-$ O $-$ Mn(Fe) bond angle ($\Theta_{B-O-B'}$) and the observed tolerance factor ($t_{f,obs}$) calculated using $d_{A-O}$ and $d_{B-O}$. Additional information extracted from the Rietveld refinement is also presented in Table S1 of the supplementary materials. According to Table 1, the highest unit cell volume ($V$) is observed for the compositions with $x = 0.15$. This is in accordance with the shift of the diffraction peaks to lower angles in this composition as it was observed in Fig.2. However, the unit cell volume decreases progressively with further increasing $Sr^{2+}$ concentration ($x > 0.15$), driven by a decrease in the average B-O bond length while the B-O-B' bond angle is slowly increasing.

In manganites, lattice distortions and the changes in structural parameters are driven by two factors: 1) the mismatch of the La (Sr)-O and Mn-O bond lengths; and 2) the presence of Jahn-Teller distortions. The impact of the sub-lattices mismatch can be better quantified using the Goldschmidt tolerance factor defined as $t_f = \frac{r_A + r_O}{\sqrt{2}(r_B + r_O)}$ [49], where $r_A$ is the average ionic radius of A-site $La^{3+}$ and $Sr^{2+}$, $r_B$ is the average ionic radius of B-site $Mn^{3+}$, $Mn^{4+}$ and $Fe^{3+}$, and $r_O$ is the ionic radius of $O^{2-}$. When $r_A$ increases while $r_B$ decreases with $x$ as seen in our case, we expect an increase in $t_f$. This tolerance factor has been well-documented for the manganites and is usually limited to the $0.75 \leq t_f \leq 1$ range [50, 51]. An orthorhombic structure is favored for $t_f < 0.96$, while a rhombohedral structure is realized for $0.96 < t_f < 1$ [51]. The observed tolerance factor determined from our Rietveld refinements can be computed using $t_{f,obs} = \frac{d_{A-O}}{\sqrt{2} d_{B-O}}$ [50], where $d_{A-O}$ and $d_{B-O}$ are determined using the refinement results. As can be seen from Table 1, the computed Goldschmidt parameter factor is close to unity and increases slightly with increasing Sr content ($x \leq 0.35$). Indeed, contrary to $Mn^{3+}$, $Mn^{4+}$ does not induce Jahn–Teller distortions and, due to its lower size and higher charge than $Mn^{3+}$, $Mn^{4+} - O^{2-}$ distances are shorter than the average $Mn^{3+} - O^{2-}$ ones. As a result, the contraction of the less distorted octahedral skeletons is leading to higher $t_{f,obs}$ values and explains the trend observed in Fig. 2 for large values of $x$.

Our observation that the rhombohedral structure is preserved over the entire composition range is different from that observed most often for bulk $La_{1-x}Sr_xMnO_3$. Manganite perovskites are usually reported to crystallize in an orthorhombic symmetry for $x$ lower than 0.17 [52]. However, according to Mitchell et al., higher symmetries (rhombohedral) can be favoured for the lowest $x$ values in $La_{1-x}Sr_xMnO_3$ ceramics if prepared in very oxidizing conditions [53]. The influence of high $Mn^{4+}$ content on symmetry was also reported for bulk $La_{1-x}Sr_xMnO_{3+\delta}$ elaborated via a soft chemistry route followed by a calcination in air at 1350°C during 6h [54]. In addition, it was observed that when prepared in air at high temperatures, $LaMnO_3$ forms the metal-vacant phase with $La_{1-\varepsilon}Mn_{1-\varepsilon}O_3$ ($\varepsilon = \frac{\delta}{(3+\delta)}$) of rhombohedral symmetry, usually described as $LaMnO_{3+\delta}$ [53,55,56]. These metal vacancies result in the oxidation of $Mn^{3+}$ into $Mn^{4+}$ in the presence of oxygen at moderate to high temperatures [53]. Thus, the persistence of the rhombohedral symmetry at our lowest $x$ values is likely a signature of metal-vacant samples leading to higher $Mn^{4+}$ content than expected from the nominal composition.

Finally, we observe in Table 1 very little changes in the unit cell lattice parameters and volume with increasing iron concentration for a fixed value of Sr content ($x$). This is consistent with the fact that $Fe^{3+}$ and $Mn^{3+}$ carry virtually identical ionic radii. Analogous weak tendencies that we have noted in our refinements have also been reported previously [50, 57-59]. A similar trend was also observed in previous works in La-Ca manganites [60-66]. To explain the slight increase in volume with the Fe content, the authors of Refs. [62,66,67] suggested the presence of a certain amount of $Fe^{4+}$ ions with an ionic radius ($r_i$ = 0.58 Å) larger than the $Mn^{4+}$ ones ($r_i$ = 0.53 Å) [45]. Our data cannot rule out this scenario although a XPS study could provide a definitive answer to the presence of these $Fe^{4+}$ ions.

From the spectra in Figs. 1(a) and (b), the average grain (domain) size of our samples can be estimated using the Debye-Sherrer formula:

$$D_{D,Sh} = \frac{K\lambda}{\beta \cos\theta} \quad (2)$$

where $K = 0.9$ is a constant, $\lambda$ is the X-ray wavelength, $\theta$ is the angular position of a selected diffraction peak and $\beta$ is its experimental full width at half-maximum (FWHM). In our case, the grain size is evaluated using the average of values computed from several diffraction peaks in the same spectra. The evolution of grain size, $D_{D,Sh}$, as a function of Sr doping is shown in Figure 5. The substitution of a larger $Sr^{2+}$ cation for $La^{3+}$ for fixed growth conditions leads to an increase of the crystallite size when $x$ increases from 0.025 to 0.15. However, $D_{D,Sh}$ decreases for Sr-rich compositions ($x > 0.15$). This trend matches that of the lattice parameters presented in Fig. 2 and in Table 1 from the Rietveld refinement fits (Table 1). A high Sr content, beyond $x = 0.15$, suppresses grain growth [46]. Such a correlation between lattice parameters, unit cell volume and nanoparticle size has already been observed [68]. It was suggested that compressive lattice strain occurs in manganite nanoparticles (due to crystallite surface tension) and becomes more important with decreasing crystallites size, because of the growing influence of their surface. We expect this grain (domain) size trend to influence the magnetic properties of our samples.

To improve the crystalline quality of our materials and to see the influence on their magnetic properties, all the samples initially pelletized at 1170°C were further annealed at various high temperatures, heated in successive steps up to 1250°C in air. To identify the most appropriate growth temperature for each composition, XRD patterns were recorded at every sintering step and their magnetic properties were also measured. XRD patterns for a succession of sintering temperatures $T_s$ for $La_{0.85}Sr_{0.15}Mn_{0.99}Fe_{0.01}O_3$ and $La_{0.85}Sr_{0.15}Mn_{0.85}Fe_{0.15}O_3$ are shown in Figure 6 (a) and (b), respectively. The patterns show a decrease in the amount of the secondary phases when increasing $T_s$. However, some extra peaks corresponding to $Mn_3O_4$ secondary phase remain in the structure even at high sintering temperature of 1250°C in $La_{0.85}Sr_{0.15}Mn_{0.99}Fe_{0.01}O_3$. As shown in Table 1 (see boldface values for $x = 0.15$, $y = 0.01$ and 0.15), the unit cell volume slightly increases when increasing the sintering temperature $T_s$. It is accompanied by a slight increase in the Mn-O bond length and a decrease in the Mn-O-Mn bond angle. This is likely the consequence of a growing density of oxygen deficiencies with sintering temperature in agreement with previous reports [69,70]. Nevertheless, the lattice parameters are evolving slowly with varying sintering conditions. Since the sintering temperature has a significant impact on the magnetic properties on many of these samples while the structural changes are minimal, other avenues like the presence of oxygen off-stoichiometry [53] or the influence of grain size and morphology must be considered to explain these changes. In what follows, we focus on grain morphology.

*Scanning electron microscopy SEM*

Figure 7 compares the grains' morphology observed by scanning electron microscopy for $La_{0.85}Sr_{0.15}Mn_{1-y}Fe_yO_3$ (y = 0.01 and 0.15) ceramics subjected to a sintering at 1070˚C [Figs. 6 (a) and (b)], 1170˚C [Figs. 6 (c) and (d)] and 1250 ˚C [Figs. 6 (e) and (f)], respectively. The images show a close-packed microstructure with grains that are clustering to form large boulders of a few microns in size. The grains have apparent sizes of approximately 500 nm for the lowest sintering temperature (1070 ˚C) but are growing beyond 1 micron in size when increasing $T_s$. Table 2 presents the average crystallite size values estimated from the SEM images ($D_{SEM}$) in Fig. 7 and that calculated from the diffraction spectra using the Debye-Sherrer formula (see Eq. 2 above). Obviously, the apparent particle sizes $D_{SEM}$ estimated from SEM are several times larger than those calculated by XRD. This indicates that each grain observed by SEM contains several smaller crystallized grains (domains) as $D_{D,Sh}$ can be envisioned as the typical domain size for coherent x-ray diffraction. These values found for $D_{D,Sh}$ agree with those observed in Ref. [71]. Although XRD and Rietveld refinement show gradual structural changes with doping and sintering temperature, we will need to consider in what follows that SEM images reveal an evolution in the microstructure that may also affect the magnetic properties of these ceramics.

*Magnetic properties*

The magnetic properties of manganites and their physical origin have been extensively studied over the last three decades [54,72-74]. Jonker and van Santen [75] and Wold and Arrott [76] independently showed that the synthesis temperature and partial oxygen pressure $P(O_2)$ can be used to control the $Mn^{3+}/Mn^{4+}$ ratio of undoped parent compound $LaMnO_3$: reducing atmosphere and/or high synthesis temperatures around 1350˚C produce samples with smaller concentrations of $Mn^{4+}$, while lower temperatures ~1100˚C and/or oxidizing atmospheres result in significant concentration of $Mn^{4+}$ affecting the magnetic properties. Of course, this $Mn^{3+}/Mn^{4+}$ ratio is also influenced by the Sr substitution for La allowing this family to exhibit for example ferromagnetism due to double exchange and related colossal magnetoresistance. Fe substitution for Mn disrupts this $Mn^{3+}/Mn^{4+}$ ratio by adding $Fe^{3+}$-O-$Mn^{3+}$ and $Fe^{3+}$-O-$Mn^{4+}$ bonds affecting the magnetic properties of these materials. In the following, we first explore the impact of these substitutions. We follow with a quick survey of the influence of the sintering temperature on the magnetic properties.

*Effect of Sr and Fe substitutions*

Figure 8 shows the field-cooled magnetization as a function of temperature in an applied magnetic field of 0.2 T for $La_{1-x}Sr_xMn_{0.99}Fe_{0.01}O_3$ in (a) and for $La_{1-x}Sr_xMn_{0.85}Fe_{0.15}O_3$ in (c), all sintered at $T_s$ = 1170°C. As shown in Fig. 8 and summarized in Table 3, the magnetization at the lowest temperature (T = 5 K) first increases with Sr substitution in the range $0.025 \leq x < 0.35$, then gradually decreases for $x \geq 0.35$. The lattice undergoes less Jahn-Teller distortions with increasing $x$ due to the reduction of the density of $Mn^{3+}$ ions, contributing to the gradual increase of the bond angle toward 180° and the increase of the tolerance factor as shown in Table 1. The evolution of the average $Mn(Fe) - O$ bond length and $Mn(Fe) - O - Mn$ bond angle upon the growing content of $Sr^{2+}$ contributes to a strengthening of the magnetic interactions while the density of ferromagnetic $Mn^{4+} - O - Mn^{3+}$ bonds is also increasing in favor of $Mn^{3+} - O - Mn^{3+}$ ones leading to ferromagnetic coupling via the double-exchange mechanism and long-range ferromagnetic order. For higher Sr contents ($x > 0.35$), the magnetization decreases. This behavior is even more pronounced for the compositions with high Fe content for which the saturation magnetization is systematically suppressed for all values of $x$.

The derivative $\frac{dM}{dT}$ as a function of T can be used to define the ferromagnetic-to-paramagnetic transition temperature $T_c$ in our samples as the inflexion point of the M (T) data as shown in Fig. 8(b) for $La_{1-x}Sr_xMn_{0.99}Fe_{0.01}O_3$ and in Fig. 8(d) for $La_{1-x}Sr_xMn_{0.85}Fe_{0.15}O_3$. The values of $T_c$ as a function of Sr content $x$ are presented in Table 3. As can be seen from Table 3, $T_c$ continuously increases with Sr content for $0.025 \leq x \leq 0.35$; y = 0.01, 0.15. For samples with higher Sr contents ($x > 0.35$), the presence of an inflexion point is less obvious from Figs. 8 (a) and (c) although the derivative curves clearly show minima. We can also note anomalies at low temperature in the derivative from the inset of Fig. 8 (b): the derivative curve for $La_{0.5}Sr_{0.5}Mn_{0.99}Fe_{0.01}O_3$ exhibits a minimum at $T_c \approx 370$ K but also a shoulder at around 250 K, while no minimum is observed within the temperature range of our measurements for $La_{0.3}Sr_{0.7}Mn_{0.99}Fe_{0.01}O_3$. We also note a similar shoulder at ~ 250 K for this latter sample indicating probably phase segregation as signaled from the analysis of the XRD patterns. In general, iron substitution for manganese leads to a strong suppression of $T_c$ but also a broadening of the transition. This is most evident for samples with $x$ = 0.35 and different Fe contents as the derivative plot gives a large peak for y = 0.15 with FWHM ~ 150 K compared to ~ 50 K for y = 0.01.

Our results for our samples with low level of iron content match well with those presented for example by Epherre and co-workers [77]. These authors showed that, for $x$ smaller than 0.25, the structural parameters and the saturation magnetization evolve slowly with $x$ while $T_c$ is continuously increasing. This low $x$ behavior is attributed to the presence of cationic vacancies in the perovskite structure resulting in a constant $Mn^{4+}$ density. From $x = 0.25$ to $0.50$, the density of vacancies at the B-site becomes small as the $Mn^{4+}$ density increases with $x$ from $\approx 35\%$ up to $\approx 50\%$ tracking closely its expected $x$ dependence [77]. Beyond $x = 0.35$, this leads to a decrease in magnetization and $T_c$ as the increasing density of $Mn^{4+}$ induces a growing competition between ferromagnetic (double exchange $Mn^{3+} - O - Mn^{4+}$) and antiferromagnetic (superexchange $Mn^{4+} - O - Mn^{4+}$) interactions. This was also shown by Hemberger *et al.* who observed a decreasing magnetization when the amount of $Mn^{4+}$ exceeded 40 % [78]. Fe substitution for Mn is adding $Fe^{3+}$-O-$Mn^{3+}$ and $Fe^{3+}$-O-$Mn^{4+}$ bonds competing with pure manganese-based bonds and thus affecting the magnetic properties of these materials. Fe doping disrupts the possibility to establish long-range magnetic order in the material, affecting in the end the magnitude of $T_c$ and leading to broad transitions.

*Effect of sintering temperature*

To tune further the magnetic and the magnetocaloric properties of our samples, we explore the impact of sintering temperature on magnetization and Curie temperature for each composition. Figure 9 shows the temperature dependence of the magnetization for $La_{1-x}Sr_xMn_{1-y}Fe_yO_3$ ($x = 0.15$, 0.5 and 0.7, $y = 0.01$ and 0.15) at a constant magnetic field of 0.2 T with the sintering temperature $T_s$ varying from 1070˚C to 1250˚C. In general, higher sintering temperature results in narrower transitions while reducing anomalies arising from secondary phases. In fact, all samples sintered at 1070˚C show an anomaly around 50 K which is constantly observed for samples prepared at low temperature, independent of $x$ and $y$, and is consistent with the presence of $Mn_3O_4$ that exhibits a magnetic phase transition around 50 K [43,44]. This feature is weakening with increasing $T_s$. A comparison between Curie temperatures of $La_{1-x}Sr_xMn_{1-y}Fe_yO_3$ ($x = 0.15, 0.5$ and $0.7, y = 0.01$ and $0.15$), sintered at 1170°C and 1250°C, extracted from the temperature dependence of $\frac{dM}{dT}$ curves at 0.2 T (Figure S2) and enlisted in Table 3, shows that contrary to $La_{1-x}Sr_xMn_{0.99}Fe_{0.01}O_3$ ($x = 0.5, 0.7$), where $T_c$ is reduced to lower temperatures when the samples were heated at 1250°C, no significant change in the minimum of the $\frac{dM}{dT}$ curves is noticed for $La_{1-x}Sr_xMn_{0.85}Fe_{0.15}O_3$ ($x = 0.5, 0.7$) compounds. In addition, as can be seen from Fig. S2, $T_c$ is clearly reduced to lower temperatures with increasing $T_s$ for $La_{0.85}Sr_{0.15}Mn_{0.85}Fe_{0.15}O_3$, while it increases with $T_s$ for $La_{0.85}Sr_{0.15}Mn_{0.99}Fe_{0.01}O_3$. Moreover, the M(T) and $\frac{dM}{dT}$ curves for $La_{0.3}Sr_{0.7}Mn_{0.99}Fe_{0.01}O_3$ sintered at 1250°C [Fig. 9(e)] clearly show two distinctive magnetic transitions at 102 K and around ~ 370 K. This low temperature transition may be related to the extra tetragonal (I4/mcm) phase observed by XRD for large Sr doping (see Fig. 2).

To better characterize the low temperature magnetization behavior of these ceramics, M (H) curves are performed at 5 K for some selected $T_s$ and are compared in Figure 10. The saturation magnetization values taken at 7 T ($M_{7T}$) for some selected samples and sintered at different temperatures are summarized in Table 3. The saturation magnetization of $La_{1-x}Sr_xMn_{0.99}Fe_{0.01}O_3$ with low Fe content is growing with $T_s$, reaching its maximum value with the maximum $T_s$ explored. This is fully consistent with previous reports showing that the magnetic, resistive and magnetoresistive properties of ceramics or polycrystalline manganites prepared by the solid-state reaction technique depend on the preparation conditions, especially on sintering and annealing temperature [79]. However, this trend is not exactly followed for samples with high Fe content as shown in Fig. 10 where the high-field magnetization is reaching a maximum at intermediate $T_s \sim$ 1170°C, matching the observations made in Fig. 9 with the temperature dependence of the magnetization. Since we do not observe a major difference in the behavior of grain size with $T_s$ for low and high Fe contents as shown in Table 2, the decrease of $T_c$ and the magnetization beyond $T_s = 1170$°C is likely affected by local compositional variations. For example, this may come from a growing density of oxygen vacancies that may have more impact when the materials are already heavily disordered by the large level of Fe content. In fact, as can also be seen from Fig. 10 (b), the decrease in the saturation magnetization of samples with large Fe content after a sintering at 1250°C is more pronounced for low $x$ ($x = 0.15$) than for large $x$ ($x = 0.5$ and $0.7$). Since $T_c$ evolves quickly with hole doping at low $x$, its strong variation with $T_s$ is consistent with an increasing density of oxygen vacancies that counters the Sr for La substitution.

Another feature of importance in Fig. 10 is that the addition of iron modifies the high field behavior of the magnetization as samples do not reach saturation even for our highest applied magnetic field and our highest explored $T_s$. This phenomenon was frequently observed in bulk manganites and was attributed to local disorder (clustering) [54, 80, 81]. This gradual increase without saturation at high fields, most noticeable with large iron content, indicates that the magnetic ground state dramatically changes from long-range to short-range ferromagnetic ordering as iron content is increased. Yusuf *et al.* [82] indicated the preservation of ferromagnetic domains up to 10% Fe doping in their Fe-doped $La_{0.67}Ca_{0.33}MnO_3$. In the same context, Barandiaràn *et al.* [83] studied $La_{0.7}Pb_{0.3}Mn_{1-x}Fe_xO_3$ $0 \leq x \leq 0.3$ and concluded that short-range ferromagnetic (FM) and antiferromagnetic (AFM) clusters of different sizes coexist in their $x = 0.2$ sample. Similarly, Barik *et al.* [32] showed the coexistence of FM and AFM clusters in $La_{0.7}Sr_{0.3}Mn_{0.8}Fe_{0.2}O_3$ with M(H) traces very similar to our data in Fig. 10 [especially Fig. 10 (f)]. Thus, Fe substitution for Mn is driving magnetic phase inhomogeneity which leads to broadened transitions, FM behavior with samples having a hard time reaching the expected saturation magnetization without sacrificing too much on the amplitude of the magnetization.

In summary, it is possible to control the magnetic properties of manganites through the usual Sr for La substitution that controls mostly the proportion of $Mn^{3+}$ and $Mn^{4+}$ ions and the dominance of the double exchange interaction in establishing the large magnetization and magnetic transition close to room temperature. Fe for Mn substitution disrupts the long-range order and drives magnetic phase inhomogeneity resulting in transition broadening and critical temperature shifts. The sintering temperature can magnify the effect of iron as it is likely leading to oxygen vacancies that adds more disorder to the system and can even affect hole doping. These three control parameters of these co-doped manganites offer an interesting avenue to tune their magnetic properties and, as will be shown below, their magnetocaloric properties in proximity to room temperature.

*Magnetocaloric properties*

The magnetocaloric effect (MCE) is an intrinsic property of magnetic materials. It is defined as the warming or the cooling of magnetic materials under the application or suppression of an external magnetic field, respectively. A goal of the present work is to explore how substitution (Sr for La, Fe for Mn) and the growth conditions ($T_s$) of a manganite-based material can be adjusted to optimize the magnitude of the isothermal magnetic entropy change ($\Delta S_M$) and the temperature range ($\Delta T_{span}$) that would allow its potential usage in cooling systems near room temperature. These parameters characterizing the MCE can be evaluated from isothermal magnetization measurements by numerically integrating the Maxwell relation found in Eq. 1 above. $\Delta S_M$ can also be determined from specific heat measurements by using the second law of thermodynamics:

$$-\Delta S_M(T, 0 \to H) = \int_0^T \frac{C_p(T',H) - C_p(T',0)}{T'} dT' \quad (3)$$

Another important parameter to determine the suitability of magnetocaloric materials for applications in cooling devices is the adiabatic temperature change $\Delta T_{ad}$. The latter can be determined from specific heat data and magnetization measurements. It is given by [1]:

$$\Delta T_{ad}(T, 0 \to H) = -\mu_0 \int_0^H \frac{T}{C_p} \left(\frac{\partial M}{\partial T}\right)_{H'} dH' \quad (4)$$

For second-order magnetic phase transitions, the adiabatic temperature change can be estimated using [1]:

$$\Delta T_{ad}(T, 0 \to H) \approx -\frac{T}{C_{p\,(H=0)}} \Delta S_M \quad (5)$$

In the following, we explore the effect of Sr/La and Fe/Mn substitutions and of the sintering temperature on the magnetocaloric effect of selected samples. For this purpose, the magnetic entropy variation $-\Delta S_M$ under several magnetic field variations of 0 to 1 T, 0 to 3 T, 0 to 5 T and 0 to 7 T is deduced using Eq. (1) from isothermal magnetization curves as those in Figure S3 of the Supplementary materials. The isothermal entropy change as a function of temperature for $La_{1-x}Sr_xMn_{1-y}Fe_yO_3$ ($x = 0.15$ and $0.35$, $y = 0.01$ and 0.15) sintered at 1170°C is presented in Figure 11. We first notice that the magnitude of $-\Delta S_M$ increases with the external magnetic field and that the maximum peak position remains nearly unaffected by the applied field for all the samples as is generally observed for other materials [1,32]. In addition, all the curves show a maximum of $-\Delta S_M$ at a temperature approaching their respective $T_c$ determined previously using the derivative of M (T) from Fig. 8.

Figs. 11 (a, c) and 11 (b, d) show that increasing the Sr content shifts the maximum peak position to higher temperatures as it tracks the evolution of $T_c$ with doping. For a fixed Sr content [comparing (a) with (b) or (c) with (d)], the peak shifts to lower temperature with increasing Fe doping. Moreover, as the magnetic inhomogeneity increases with Fe content, the maximum value of $-\Delta S_M$ decreases but the peak widens over a larger temperature range around $T_c$. This behavior is in accordance with those obtained by Barik *et al.* [32] and can be mainly attributed, as mentioned previously, to the suppression of the long-range ferromagnetic order as many of the $Mn^{4+}$-O- $Mn^{3+}$ DE bonds are replaced by a large number of antiferromagnetic SE bonds between $Mn^{3+}$ and $Fe^{3+}$ competing with ferromagnetic ones between $Mn^{4+}$ and $Fe^{3+}$ as was observed in $La_2MnFeO_6$ and $LaSrMnFeO_6$ [84]. Thus, it is possible to shift the maximum in $-\Delta S_M(T)$ close to room temperature with a wise choice of Sr and Fe concentrations and control the width of the $-\Delta S_M(T)$ peak (defined here as $\Delta T_{span}$) over which it remains important. In some cases, $\Delta T_{span}$ extends way over 150 K [see Figs. 11 (a) and (d) for $x = 0.15$, $y = 0.01$ and $x = 0.35$, $y = 0.15$, respectively].

Figs.11 (e) and (f) present examples of the effect of the sintering temperature on the magnetocaloric effect of our samples. The magnetic entropy variation for the $La_{0.85}Sr_{0.15}Mn_{0.99}Fe_{0.01}O_3$ and $La_{0.85}Sr_{0.15}Mn_{0.85}Fe_{0.15}O_3$ ceramics sintered at 1250°C under several magnetic field variations of 0 to 1 T, 0 to 3 T, 0 to 5 T and 0 to 7 T shows that the maximum peak position of $-\Delta S_M$ for $La_{0.85}Sr_{0.15}Mn_{0.99}Fe_{0.01}O_3$ remains nearly field independent even after sintering [Fig. 11 (e)]. In addition, the magnitude of $-\Delta S_M$ reaches 4.7 J/kg K for a magnetic field variation of 0 to 7 T compared to 3.0 J/kg K for the sample sintered at 1170°C [see Fig. 11(a)]. This increase of $-\Delta S_M$ with $T_s$ is consistent with the increase of the saturation magnetization as a function of $T_s$ observed in Fig. 10 (a). Comparing further the samples in Figs.11 (a) and (e) differing only by the sintering temperature, the $-\Delta S_M$ peaks of the sample prepared at 1250°C become narrower compared to that sintered at 1170°C. This indicates that sintering temperature can also be used as a tool to control the amount of magnetic inhomogeneities in the samples as in the case of Fe doping.

Furthermore, the impact of sintering at higher temperature has the opposite effect for samples with large Fe substitution levels. This is shown for example with $La_{0.85}Sr_{0.15}Mn_{0.85}Fe_{0.15}O_3$ for which the temperature of maximum entropy change at 7T shifts from 175 down to 102 K for $T_s$ varying from 1170 to 1250°C. This reduction in the maximum $-\Delta S_M$ temperature is also accompanied by a broadening of the temperature range. Again, this trend correlates well with the $T_c$ shift observed in Fig. 9 (b) and the decrease in magnetization reported in Figs. 10 (b).

Altogether, the magnetocaloric effect is sensitive to the actual proportions of Sr for La and Fe for Mn substitutions that play into the doping to adjust the strength and dominance of ferromagnetic coupling, but also using disorder as a tool to broaden and adjust the temperature range with significant magnetic entropy change. Our data show that an appropriate choice for both can be used to optimize the isothermal entropy change for a given (target) temperature range that requires controlling the temperature of the maximum $-\Delta S_M$ but also the temperature range ($\Delta T_{span}$) over which it is significant. Finally, the sintering temperature can also be used to tune the magnetocaloric properties.

Using specific heat data measured at 0 T (Figure 12) and the isothermal magnetic entropy changes [Figs. 11 (a) and (c)], the adiabatic temperature change as a function of temperature for $La_{0.85}Sr_{0.15}Mn_{0.99}Fe_{0.01}O_3$ and $La_{0.65}Sr_{0.35}Mn_{0.99}Fe_{0.01}O_3$ is calculated using Eq.(5) and is shown in Figures 13 (a) and (b), respectively. As expected for both samples, $\Delta T_{ad}$ shows a maximum at $T_c$. It reaches 3 K for $La_{0.85}Sr_{0.15}Mn_{0.99}Fe_{0.01}O_3$ and 2.9 K for $La_{0.65}Sr_{0.35}Mn_{0.99}Fe_{0.01}O_3$ for a magnetic field change of 7T. Additional Fe substitution suppresses $\Delta T_{ad}$ roughly by a factor of 2 as a result of the decreasing magnitude of $-\Delta S_M$ (see Fig. 11) and assuming the same magnitude for the specific heat. For both $La_{0.85}Sr_{0.15}Mn_{0.99}Fe_{0.01}O_3$ and $La_{0.65}Sr_{0.35}Mn_{0.99}Fe_{0.01}O_3$, adiabatic temperature changes remain moderate when compared to reference magnetocaloric materials [1]. This can be explained essentially by their low entropy changes compared to other materials but also by their large specific heat dominated by the phonon contribution.

To achieve MCE performances suitable to applications, close to room temperature, a large $(-\Delta S_{M,max})$ over a wide temperature span is strongly recommended [1,84]. To explore the magnetocaloric performance of our magnetic refrigerants, we have calculated the relative cooling power (RCP) as it allows one to compare the cooling performances of different materials. It considers the magnitude of $-\Delta S_M$, but also the temperature range $\Delta T_{span}$ for which it remains significant. It is defined as the product of the maximum value $(-\Delta S_{M,max})$ and the full width at half-maximum ($\delta T_{FWHM}$) of the peaks observed in Fig. 11 [31]:

$$\text{RCP (S)} = -\Delta S_{M,max} \times \delta T_{FWHM} \quad (6)$$

Figure 14 (a) presents the RCP at 7 T as a function of Sr content for $La_{1-x}Sr_xMn_{0.99}Fe_{0.01}O_3$ ($x \leq 0.35$) sintered at 1170°C. For comparison, the maximum entropy change $(-\Delta S_{M,max})$ as a function of Sr content is also presented. The relative cooling power (RCP) values at 7 T are found to vary between 460 and 390 J/kg, comparing well with other oxides [85-87]. Despite the increase of $-\Delta S_{M,max}$ with increasing Sr content, the RCP decreases. In fact, as shown in Figure 14 (b), it is directly related to a decrease of the full width at half-maximum ($\delta T_{FWHM}$) as $x$ increases. These results emphasize the fact that the best doping for the highest RCP is not that corresponding to the maximum $T_c$ ($x = 0.35$), but rather a compromise at $x \sim 0.2$ that leads to a large enough entropy change at room temperature and a $-\Delta S_M$ peak broadened by magnetic phase inhomogeneity. This highlights the importance of extending the working temperature range on the performance of magnetic refrigerants and justifies also using Fe for Mn substitution to tune further these performances.

Our results demonstrate that compounds with relatively high $-\Delta S_M$, but not necessarily the largest ones, and large RCP values due to a large temperature range of significant $-\Delta S_M$, can be synthesized. Their exact properties can be controlled mostly by Sr for La, Fe for Mn substitutions and by the growth conditions, leading to imperfect samples with broad transitions that could be nevertheless of interest for applications in room-temperature magnetocaloric devices. Altogether, we see that the ferromagnetic properties of these co-doped manganites can be adjusted. We can use Sr and Fe substitution to control the actual $T_c$ of the samples and the magnitude of the magnetization. These substitutions affect their magnetization field dependence and the broadness of the transition, controlled by the presence of magnetic phase segregation. The choice of sintering temperature is another lever one can use to finely tune the properties with the goal of maximizing the magnetocaloric effect in a given temperature window.

We should underline that the MCE of these ceramics remains moderate despite all our manipulations. As was shown previously, larger $-\Delta S_M$ can be achieved in manganites by substituting Ca for Sr in $La_{2/3}(Ca_{1-x}Sr_x)_{1/3}MnO_3$ [88]. As the crystal symmetry changes to Pnma for Ca-rich compositions (for x < 0.15), $-\Delta S_M$ is also magnified while the transition temperature is decreasing [88]. This Ca for La substitution path was explored previously by our group in Ref. [84] as we substituted Ca for La into $La_2MnFeO_6$ (LMFO). Contrary to Ca-substituted $(La,Sr)MnO_3$, Ca-doped LMFO shows poor ferromagnetism (weak magnetization) and weak MCE despite observing the same transition in crystal symmetry. We concluded in Ref. [84] that a very small B-O-B' bond angle was at the origin of the weak magnetic interaction, together with cation disorder. The same decrease in bond angle is also observed in $(La,Ca)MnO_3$, explaining the suppression of the optimal $T_c$. We note however that there may be some interest to look for the same gradual Fe substitution for Mn we have been exploring in this paper into $La_{2/3}(Ca_{1-x}Sr_x)_{1/3}MnO_3$ as a source of disordering that could broaden the transition while taking advantage of the increase in MCE.

**Conclusion**

In summary, we have investigated the structural, magnetic and magnetocaloric properties of $La_{1-x}Sr_xMn_{1-y}Fe_yO_3$ $(0.025 \leq x \leq 0.7; y = 0.01, 0.15)$ perovskite manganite compounds. We show how one can tune the magnetic and the magnetocaloric properties of these manganite perovskite oxides by chemical substitution and/or growth conditions. We show also that Sr substitution for La favors mainly double-exchange interaction leading to higher magnetization and $T_c$ values, while Fe substitution for Mn drives magnetic disorder. Sintering temperature is another tool to control the magnetic disorder.

All the ceramic samples crystallize in a rhombohedral structure ($R\bar{3}c$) in a large proportion with a decrease of the unit cell volume as Sr content increases. The temperature dependence of the magnetization shows a macroscopic ferromagnetic-like behavior for all compounds. The magnetic and magnetocaloric properties are strongly affected by the chemical substitution and the sintering temperature. Our data reveals that the maximum magnetic entropy change $(-\Delta S_{M,max})$ at $T_c$ continuously increases with Sr content up to $x \sim 0.35$ and decreases for larger substitution levels. Fe for Mn substitution suppresses the magnitude of $-\Delta S_{M,max}$, shifts down the transition temperature, but leads also to a broaden temperature range $\Delta T_{span}$ with large magnetic entropy change. This operating temperature range is thus affected by the Sr and Fe contents and the sintering temperature. In this way, a significant entropy change over a broad temperature range can be obtained around room temperature. Due to their relatively high magnetic entropy changes, large operating temperature range and high RCP values, the Sr doped manganite perovskite samples with properties fine-tuned by Fe substitution for Mn could be of interest for applications in magnetocaloric devices at room temperature. With the appropriate control of their stoichiometry through chemical substitution and their exact growth conditions, one can tune their magnetocaloric in a targeted range of temperature for specific cooling applications.

**ACKNOWLEDGMENTS**

The authors thank M. Castonguay, S. Pelletier, B. Rivard and M. Dion for technical support. M. Balli acknowledges funding by the International University of Rabat, Morocco. This work is supported by the Natural Sciences and Engineering Research Council of Canada (NSERC) under grant RGPIN-2018-06656, the Canada First Research Excellence Fund (CFREF), the Fonds de Recherche du Québec - Nature et Technologies (FRQNT) and the Université de Sherbrooke.

**Tables**

| Fe content (y) | y = 0.01 | | | | | y = 0.15 | | | | |
|---|---|---|---|---|---|---|---|---|---|---|
| Sr content (x) | 0.025 | 0.15 | 0.35 | 0.5 | 0.7 | 0.025 | 0.15 | 0.35 | 0.5 | 0.7 |
| Space group | R-3c | | | | | R-3c | | | | |
| $a$ (Å) | 5.531 | 5.535 / **5.536** | 5.487 | 5.529 | 5.488 | 5.524 | 5.532 / **5.533** | 5.499 | 5.460 | 5.456 |
| $c$ (Å) | 13.345 | 13.369 / **13.369** | 13.348 | 13.346 | 13.344 | 13.431 | 13.469 / **13.483** | 13.36 | 13.475 | 13.46 |
| $V$ (Å³) | 353.61 | 354.84 / **354.97** | 348.10 | 353.32 | 348.10 | 355.03 | 357.05 / **357.47** | 349.93 | 348.01 | 347.12 |
| $d_{A-O}$ (Å) | 2.752 | 2.751 / **2.755** | 2.734 | 2.756 | 2.748 | 2.765 | 2.763 / **2.777** | 2.742 | 2.749 | 2.752 |
| $d_{B-O}$ (Å) | 1.965 | 1.963 / **1.967** | 1.939 | 1.971 | 1.960 | 1.972 | 1.967 / **1.981** | 1.949 | 1.944 | 1.950 |
| $\Theta_{B-O-B}$ (°) | 163.82 | 164.42 / **164.15** | 172.540 | 161.31 | 161.8 | 162.24 | 163.86 / **160.59** | 168.28 | 169.27 | 165.02 |
| $t_{f,obs}$ | 0.990(0) | 0.991(0) / **0.988(2)** | 0.996(6) | 0.988(6) | 0.991(3) | 0.991(4) | 0.995(2) / **0.991(2)** | 0.994(8) | 0.999(9) | 0.998(0) |

**Table 1**: Crystal structure parameters extracted from the Rietveld refinements. It includes the lattice parameters (a and c) and unit cell volume (V), the average La (Sr)-O distance ($d_{A-O}$), the average Mn (Fe)-O bond length ($d_{B-O}$), the average Mn (Fe)-O-Mn bond angle ($\Theta_{B-O-B}$) and the observed tolerance factor ($t_{f,obs}$). All the data are for samples grown at 1170°C, except for the boldface ones (x = 0.15, y = 0.01 and 0.15) that are additionally sintered at 1250°C.

| Compounds | $D_{SEM}$ (µm) | | | $D_{D,Sh}$ (nm) | | |
|---|---|---|---|---|---|---|
| | 1070°C | 1170°C | 1250°C | 1070°C | 1170°C | 1250°C |
| $La_{0.85}Sr_{0.15}Mn_{0.99}Fe_{0.01}O_3$ | 0.6 | 1.6 | 2.5 | 19.9 | 20.3 | 20.6 |
| $La_{0.85}Sr_{0.15}Mn_{0.85}Fe_{0.15}O_3$ | 0.7 | 1.5 | 2.2 | 20.8 | 22.4 | 23.2 |

**Table 2:** Comparison between average grain sizes extracted from XRD patterns and SEM images.

|  | y = 0.01 | | | | | | y = 0.15 | | | | | |
|---|---|---|---|---|---|---|---|---|---|---|---|---|
| T$_s$ (°C) | 1170 | | | 1250 | | | 1170 | | | 1250 | | |
| Compounds | T$_c$ (K) | M at 5 K ($\mu_B$/f.u.) 0.2 T | M$_{7T}$ ($\mu_B$/f.u.) | T$_c$ (K) | M at 5 K ($\mu_B$/f.u.) 0.2 T | M$_{7T}$ ($\mu_B$/f.u.) | T$_c$ (K) | M at 5 K ($\mu_B$/f.u.) 0.2 T | M$_{7T}$ ($\mu_B$/f.u.) | T$_c$ (K) | M at 5 K ($\mu_B$/f.u.) 0.2 T | M$_{7T}$ ($\mu_B$/f.u.) |
| La$_{0.997}$Sr$_{0.025}$Mn$_{1-y}$Fe$_y$O$_3$ | 142 | 2.4 | 3.6 | - | - | - | 102 | 1.58 | - | - | - | - |
| La$_{0.85}$Sr$_{0.15}$Mn$_{1-y}$Fe$_y$O$_3$ | 255 | 3 | 3.55 | 261 | 2.83 | 3.88 | 161 | 2.08 | 2.7 | 91 | 0.44 | 0.9 |
| La$_{0.65}$Sr$_{0.35}$Mn$_{1-y}$Fe$_y$O$_3$ | 374.4 | 2.8 | 3.5 | - | - | - | 212.5 | 2.0 | 2.8 | - | - | - |
| La$_{0.50}$Sr$_{0.50}$Mn$_{1-y}$Fe$_y$O$_3$ | 371 | 2.03 | 2.60 | 351 | 2.08 | 2.70 | 252 | 1.53 | 2.16 | 252 | 1.43 | 2.0 |
| La$_{0.30}$Sr$_{0.70}$Mn$_{1-y}$Fe$_y$O$_3$ | - | 1.34 | 1.85 | 371 | 1.38 | 2.05 | 251 | 0.48 | 0.9 | 251 | 0.4 | 0.8 |

**Table 3**: Transition temperatures, low temperature magnetization (5K), saturation magnetization taken at 7T for La$_{1-x}$Sr$_x$Mn$_{1-y}$Fe$_y$O$_3$ samples sintered at 1170 °C and at 1250 °C.

**FIGURE CAPTIONS**

**Figure 1:** Powder XRD patterns of $La_{1-x}Sr_xMn_{1-y}Fe_yO_3$ ($0.025 \leq x \leq 0.7$) compounds prepared at $T_s = 1170°C$ for y = 0.01 in **(a)** and y = 0.15 in **(b)**. Secondary phases are identified as follows: ♦ for $Mn_3O_4$, ♠ for $SrCO_3$ and ▽ for $La_2O_3$.

**Figure 2:** Magnified view of the XRD peak with the highest intensity ($2\theta \approx 32°$) for $La_{1-x}Sr_xMn_{1-y}Fe_yO_3$ ($0.025 \leq x \leq 0.7$) prepared at $T_s = 1170°C$ for y = 0.01 in **(a)** and y = 0.15 in **(b)**.

**Figure 3:** Powder XRD patterns and Rietveld refinement fits of $La_{0.975}Sr_{0.025}Mn_{1-y}Fe_yO_3$ compounds prepared at $T_s = 1170°C$ for y = 0.01 in **(a)** and y = 0.15 in **(b)**. The refinement fits include the possible presence of various manganite symmetries and of $Mn_3O_4$.

**Figure 4:** Phase fractions as a function of nominal strontium doping level in the $La_{1-x}Sr_xMn_{0.99}Fe_{0.01}O_3$ ($0.025 \leq x \leq 0.7$) samples sintered at 1170°C.

**Figure 5:** Crystallites size from the Debye-Sherrer equation as a function of Sr content in $La_{1-x}Sr_xMn_{1-y}Fe_yO_3$ ($0.025 \leq x \leq 0.7, y = 0.01, 0.15$).

**Figure 6:** Powder XRD patterns for **(a)** $La_{0.85}Sr_{0.15}Mn_{0.99}Fe_{0.01}O_3$ and **(b)** for $La_{0.85}Sr_{0.15}Mn_{0.85}Fe_{0.15}O_3$ samples prepared with different final sintering temperatures $T_s$. Secondary phases are observed in a few samples and are identified as follows: ♦ for $Mn_3O_4$, ♠ for $SrCO_3$ and ▽ for $La_2O_3$.

**Figure 7:** SEM images for $La_{0.85}Sr_{0.15}Mn_{1-y}Fe_yO_3$ (y = 0.01 and 0.15, respectively), sintered at **(a and b)** 1070°C, **(c and d)** 1170°C, **(e and f)** 1250°C.

**Figure 8:** Magnetization as a function of temperature for **(a)** $La_{1-x}Sr_xMn_{0.99}Fe_{0.01}O_3$ ($0.025 \leq x \leq 0.7$) and **(c)** $La_{1-x}Sr_xMn_{0.85}Fe_{0.15}O_3$ ($0.025 \leq x \leq 0.7$) samples sintered at $T_s = 1170°C$ under an applied magnetic field of 0.2 T. The derivative $\frac{dM}{dT}$ as a function of T for **(b)** $La_{1-x}Sr_xMn_{0.99}Fe_{0.01}O_3$ ($0.025 \leq x \leq 0.7$) and **(d)** $La_{1-x}Sr_xMn_{0.85}Fe_{0.15}O_3$ ($0.025 \leq x \leq 0.7$) samples. Inset in (b) is for x = 0.5 and 0.7 while inset in (d) is for x = 0.7.

**Figure 9:** Magnetization as a function of temperature for various sintering temperature $T_s$ for **(a)** $La_{0.85}Sr_{0.15}Mn_{0.99}Fe_{0.01}O_3$, **(b)** $La_{0.85}Sr_{0.15}Mn_{0.85}Fe_{0.15}O_3$, **(c)** $La_{0.5}Sr_{0.5}Mn_{0.99}Fe_{0.01}O_3$, **(d)** $La_{0.5}Sr_{0.5}Mn_{0.85}Fe_{0.15}O_3$, **(e)** $La_{0.3}Sr_{0.7}Mn_{0.99}Fe_{0.01}O_3$ and **(f)** $La_{0.3}Sr_{0.7}Mn_{0.85}Fe_{0.15}O_3$.

**Figure 10:** Magnetization as a function of magnetic field at 5 K for various sintering temperature $T_s$ for **(a)** $La_{0.85}Sr_{0.15}Mn_{0.99}Fe_{0.01}O_3$, **(b)** $La_{0.85}Sr_{0.15}Mn_{0.85}Fe_{0.15}O_3$, **(c)** $La_{0.5}Sr_{0.5}Mn_{0.99}Fe_{0.01}O_3$, **(d)** $La_{0.5}Sr_{0.5}Mn_{0.85}Fe_{0.15}O_3$, **(e)** $La_{0.3}Sr_{0.7}Mn_{0.99}Fe_{0.01}O_3$ and **(f)** $La_{0.3}Sr_{0.7}Mn_{0.85}Fe_{0.15}O_3$.

**Figure 11:** Temperature dependence of the magnetic entropy change under different magnetic field variations for **(a)** $La_{0.85}Sr_{0.15}Mn_{0.99}Fe_{0.01}O_3$, **(b)** $La_{0.85}Sr_{0.15}Mn_{0.85}Fe_{0.15}O_3$, **(c)** $La_{0.65}Sr_{0.35}Mn_{0.99}Fe_{0.01}O_3$ and **(d)** $La_{0.65}Sr_{0.35}Mn_{0.85}Fe_{0.15}O_3$ and for **(e)** $La_{0.85}Sr_{0.15}Mn_{0.99}Fe_{0.01}O_3$, **(f)** $La_{0.85}Sr_{0.15}Mn_{0.85}Fe_{0.15}O_3$. **(a) – (d)**: samples sintered at 1170˚C , **(e)** and **(f)** : samples sintered at 1250˚C.

**Figure 12:** Specific heat as a function of temperature in zero magnetic field for $La_{0.85}Sr_{0.15}Mn_{0.99}Fe_{0.01}O_3$ and $La_{0.65}Sr_{0.35}Mn_{0.99}Fe_{0.01}O_3$.

**Figure 13:** Temperature dependence of the adiabatic temperature change $\Delta T_{ad}$ for **(a)** $La_{0.85}Sr_{0.15}Mn_{0.99}Fe_{0.01}O_3$ and **(b)** $La_{0.65}Sr_{0.35}Mn_{0.99}Fe_{0.01}O_3$ as a function of temperature.

**Figure 14:** Relative cooling power (RCP) and maximum magnetic entropy change as a function of the strontium content in **(a)** Tc and full width at half maximum as a function of the Sr content in **(b)**.

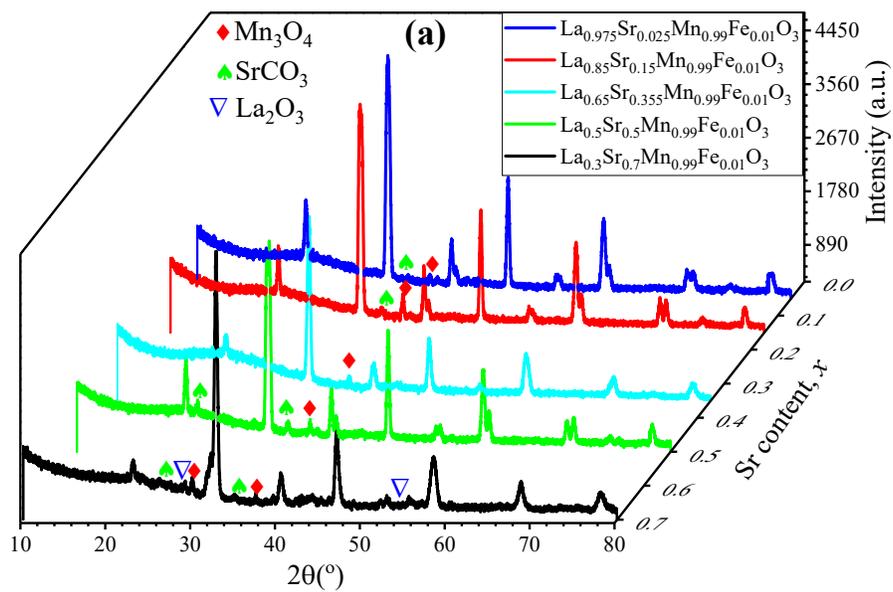

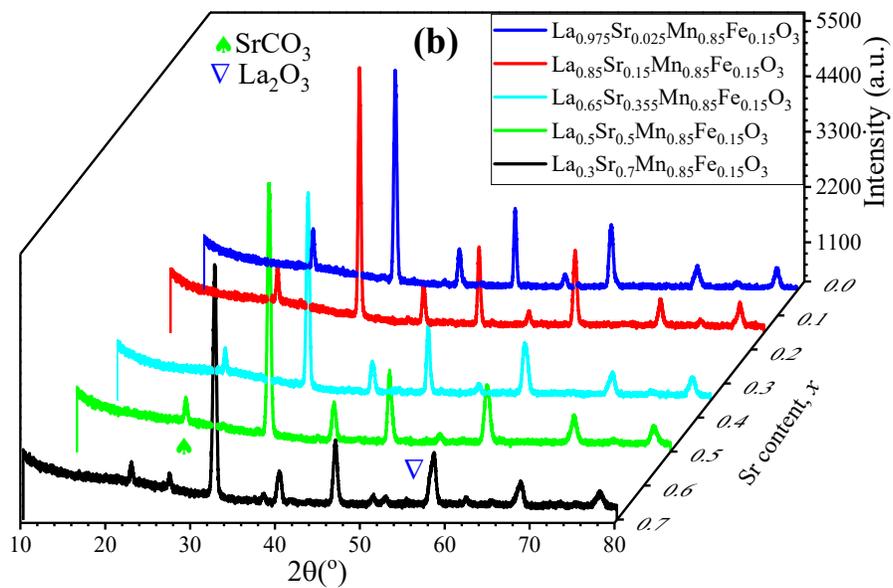

**Figure 1**

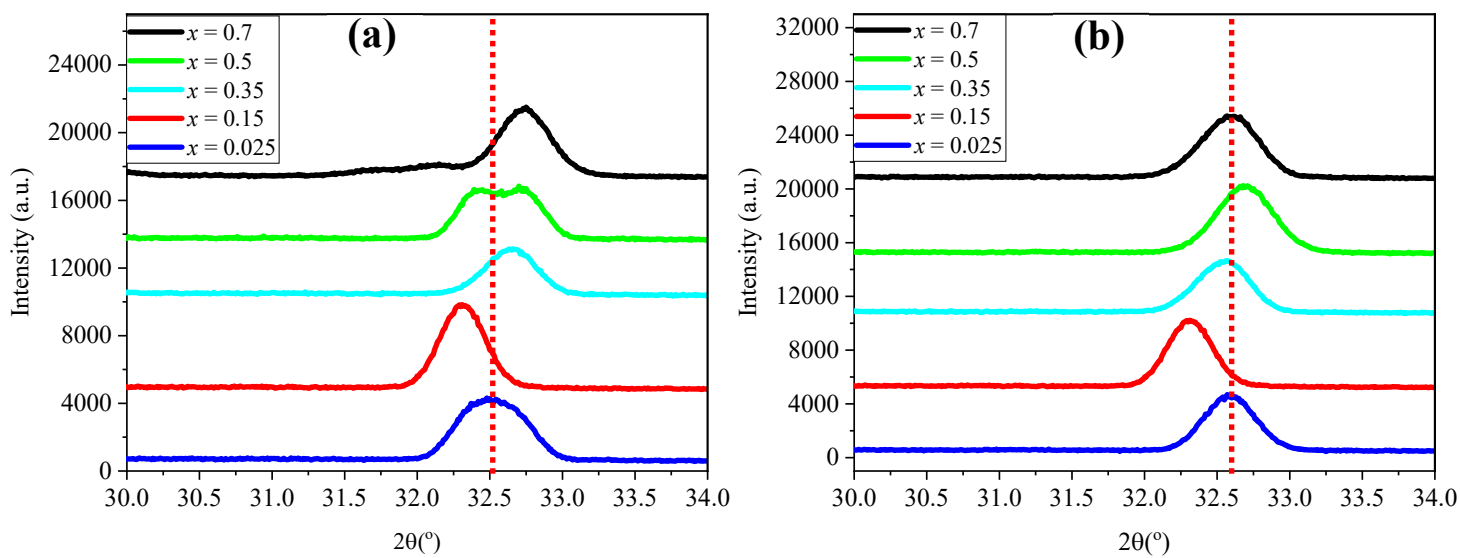

**Figure 2**

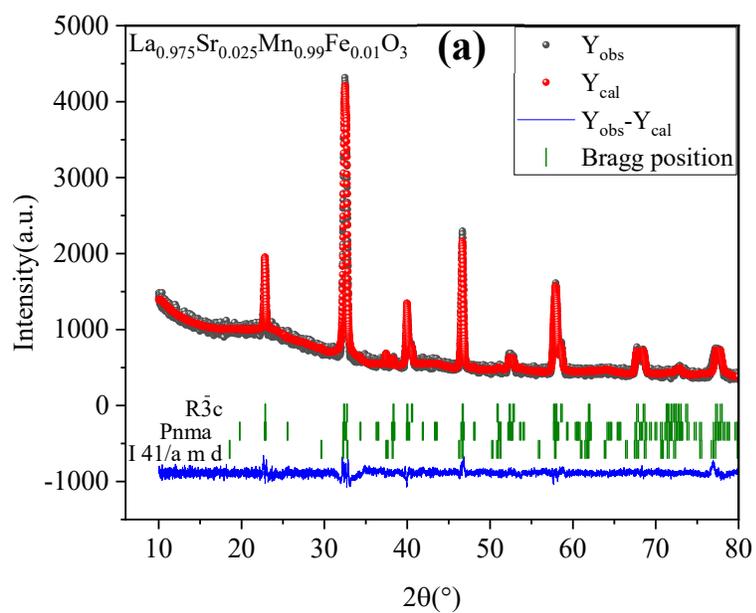

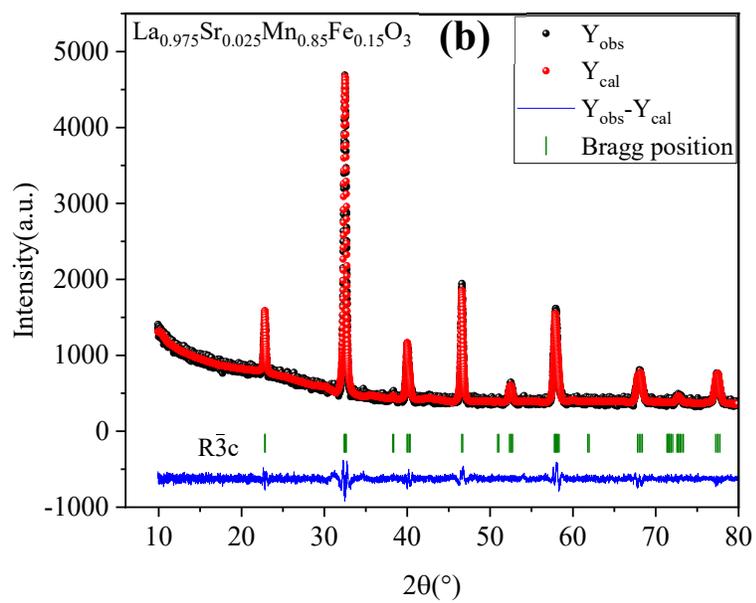

**Figure 3**

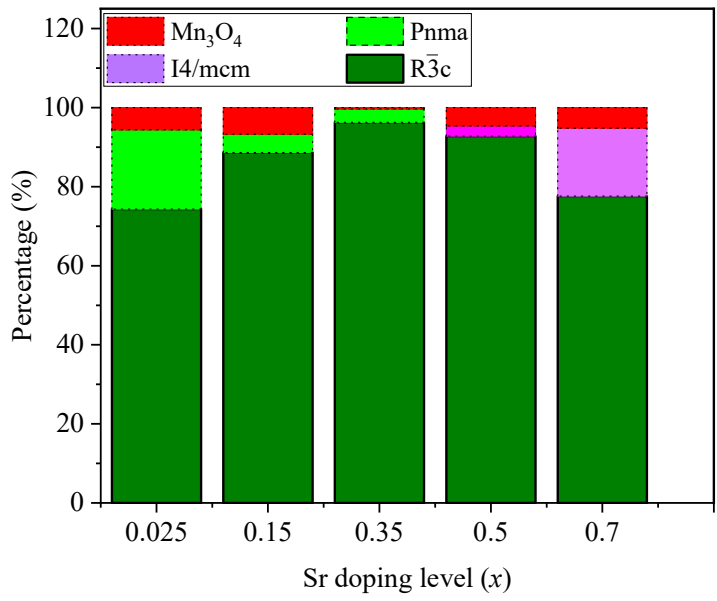

**Figure 4**

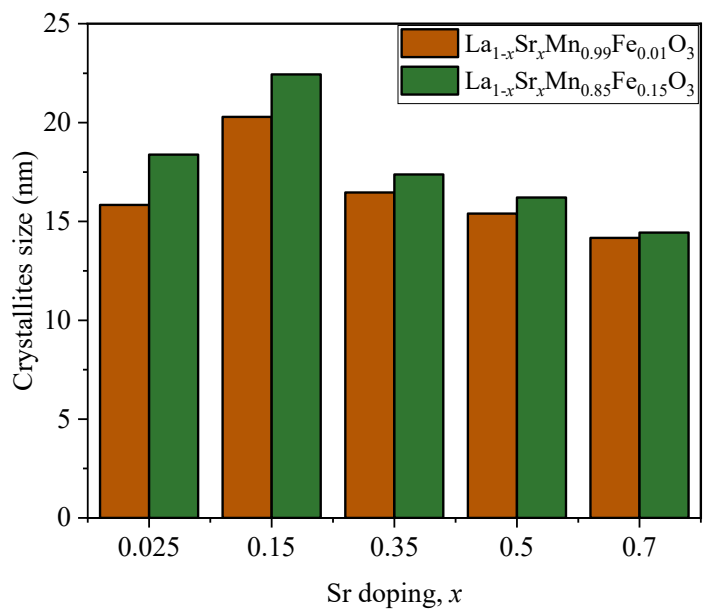

**Figure 5**

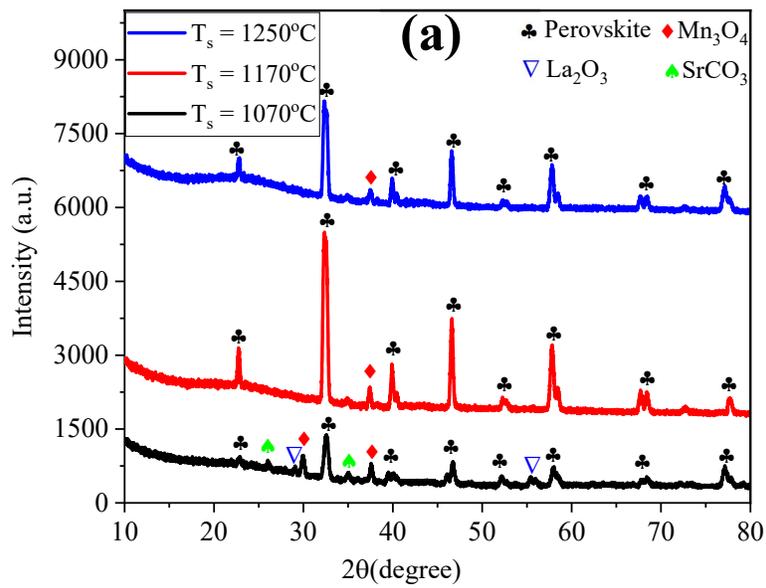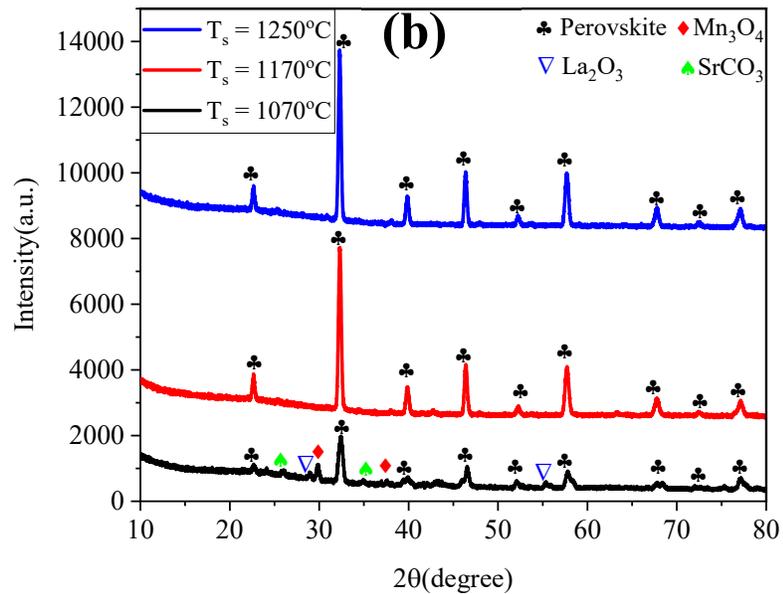

**Figure 6**

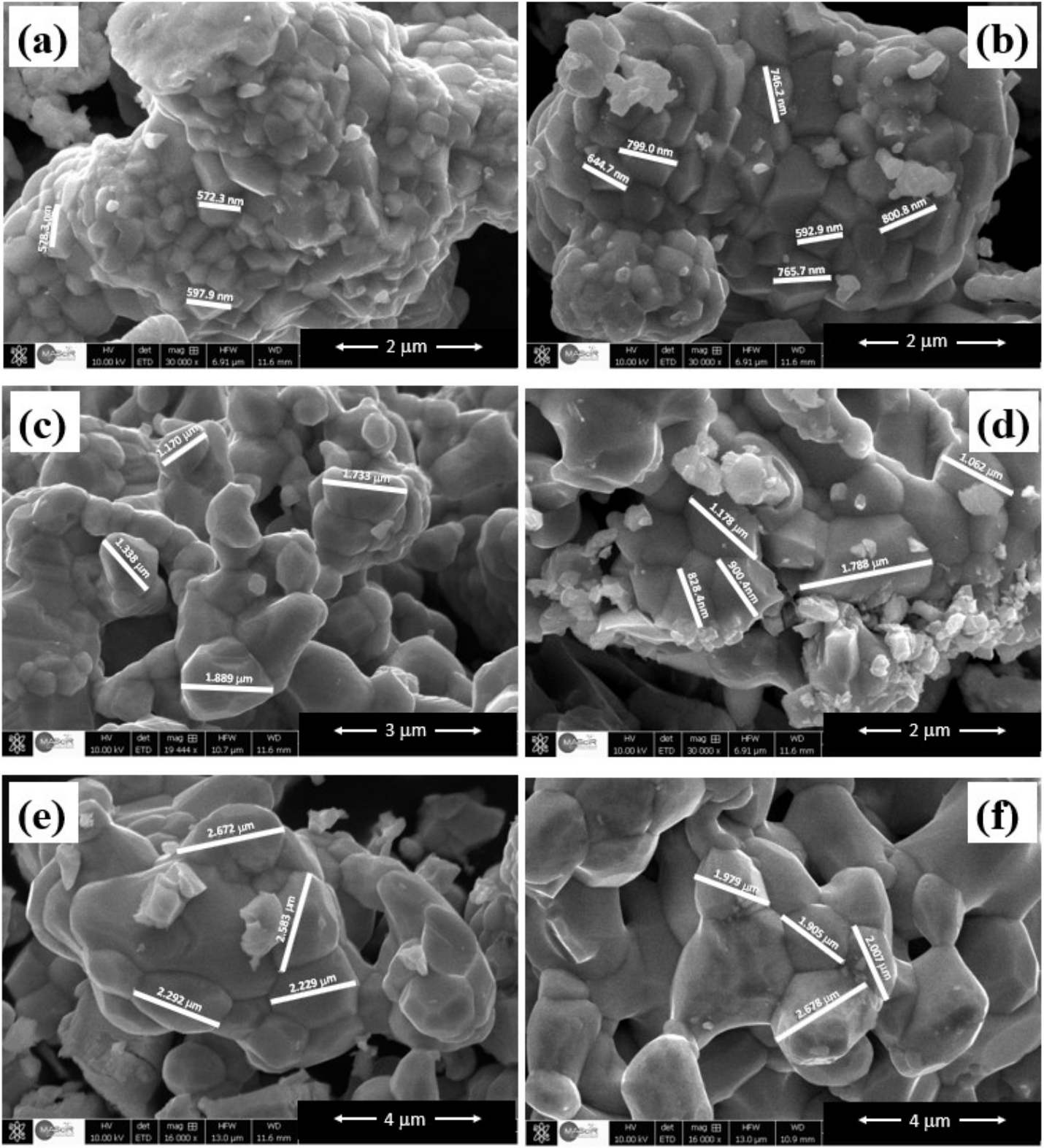

**Figure 7**

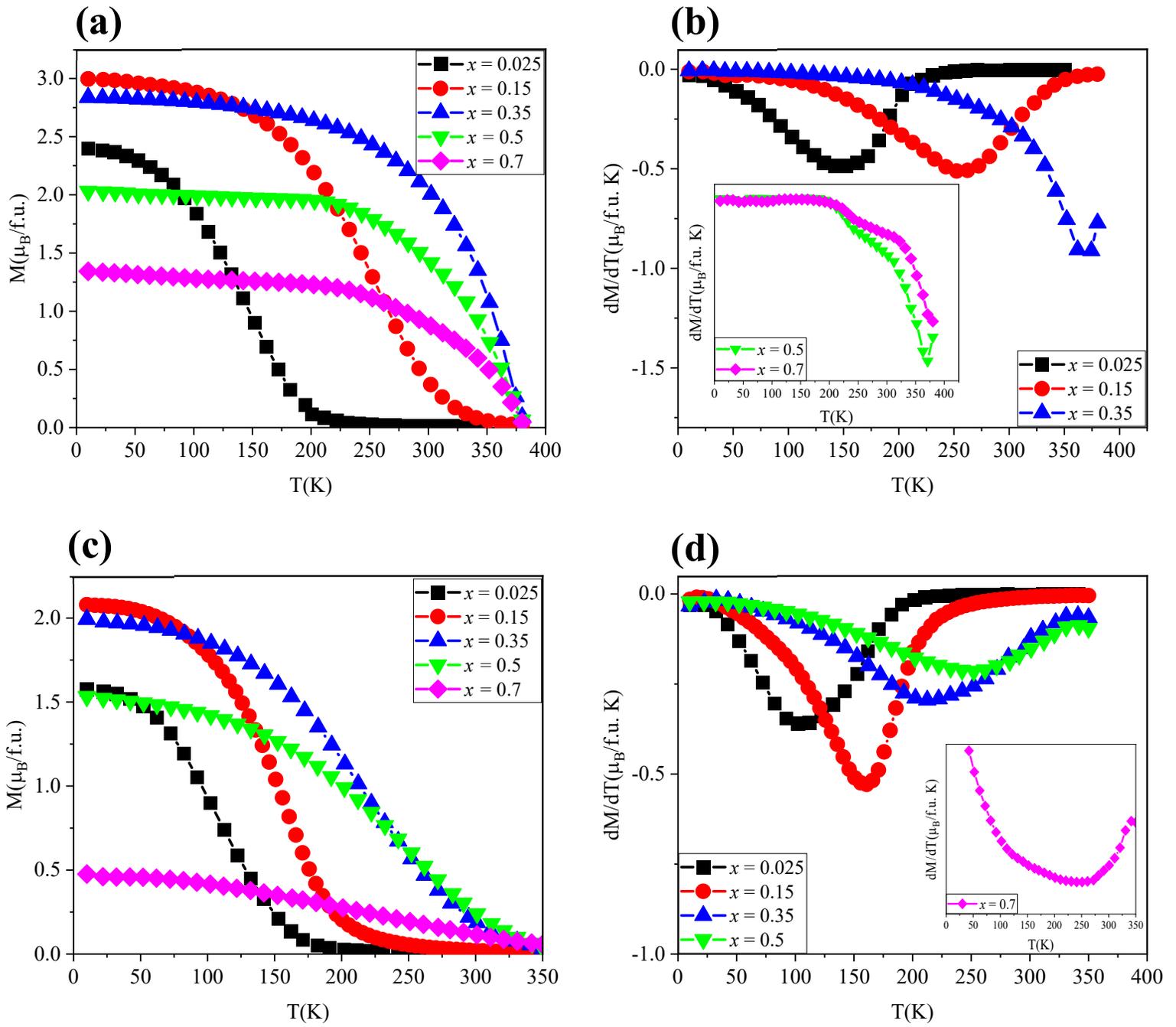

Figure 8

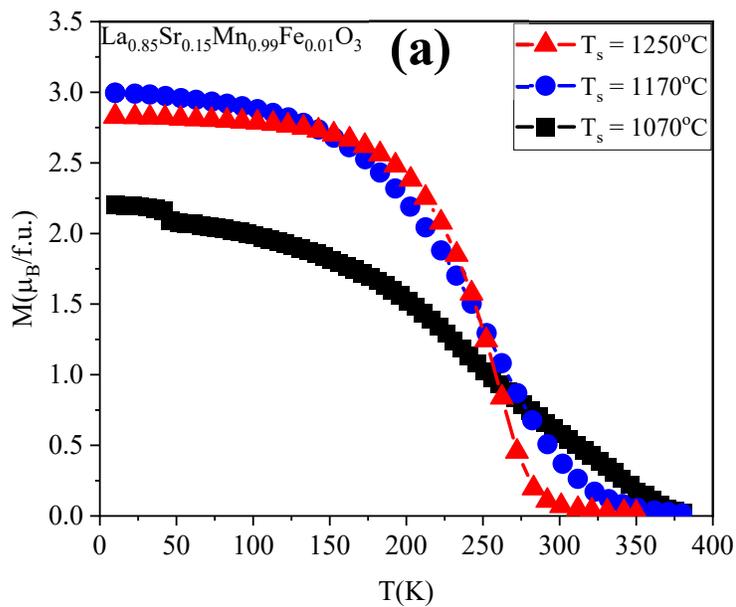
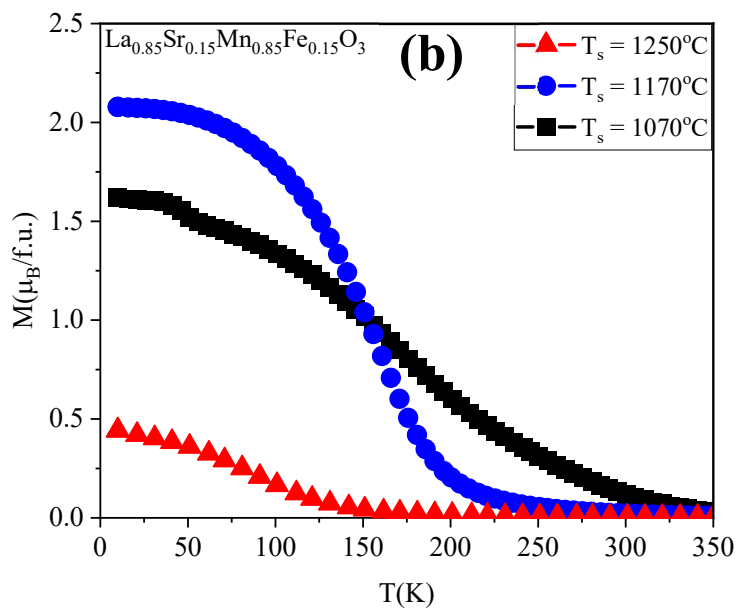
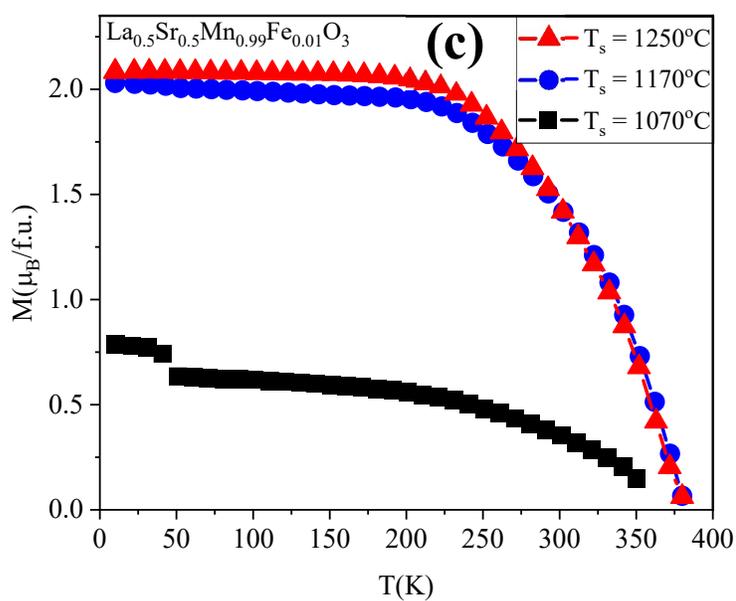
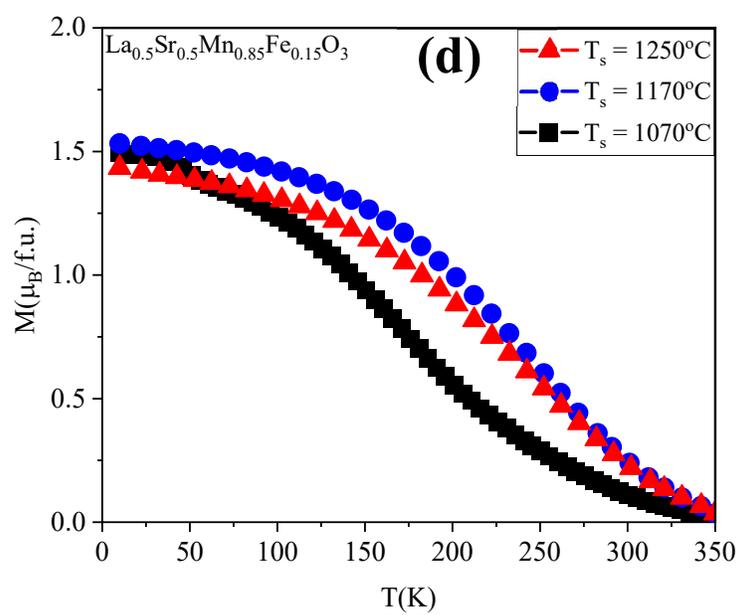
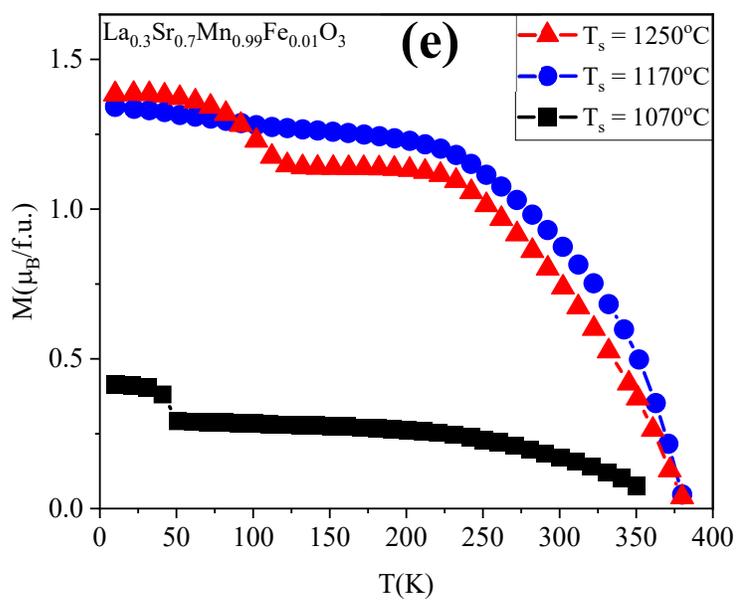
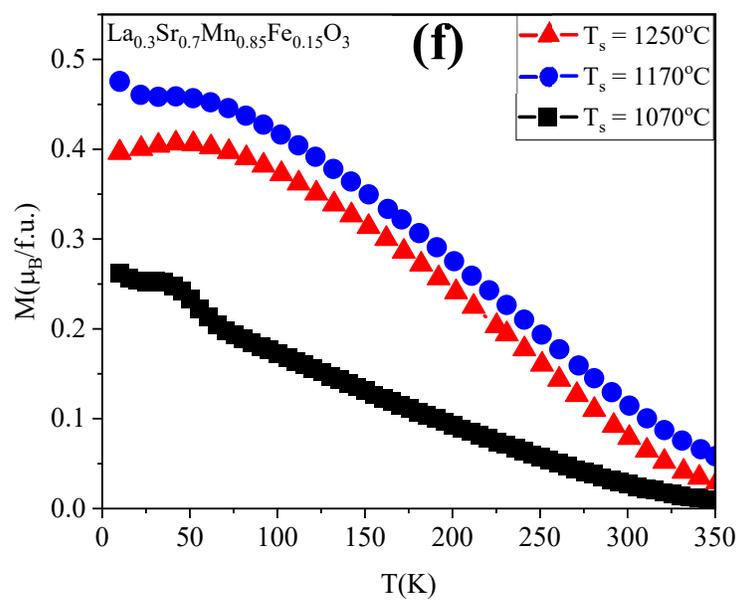

**Figure 9**

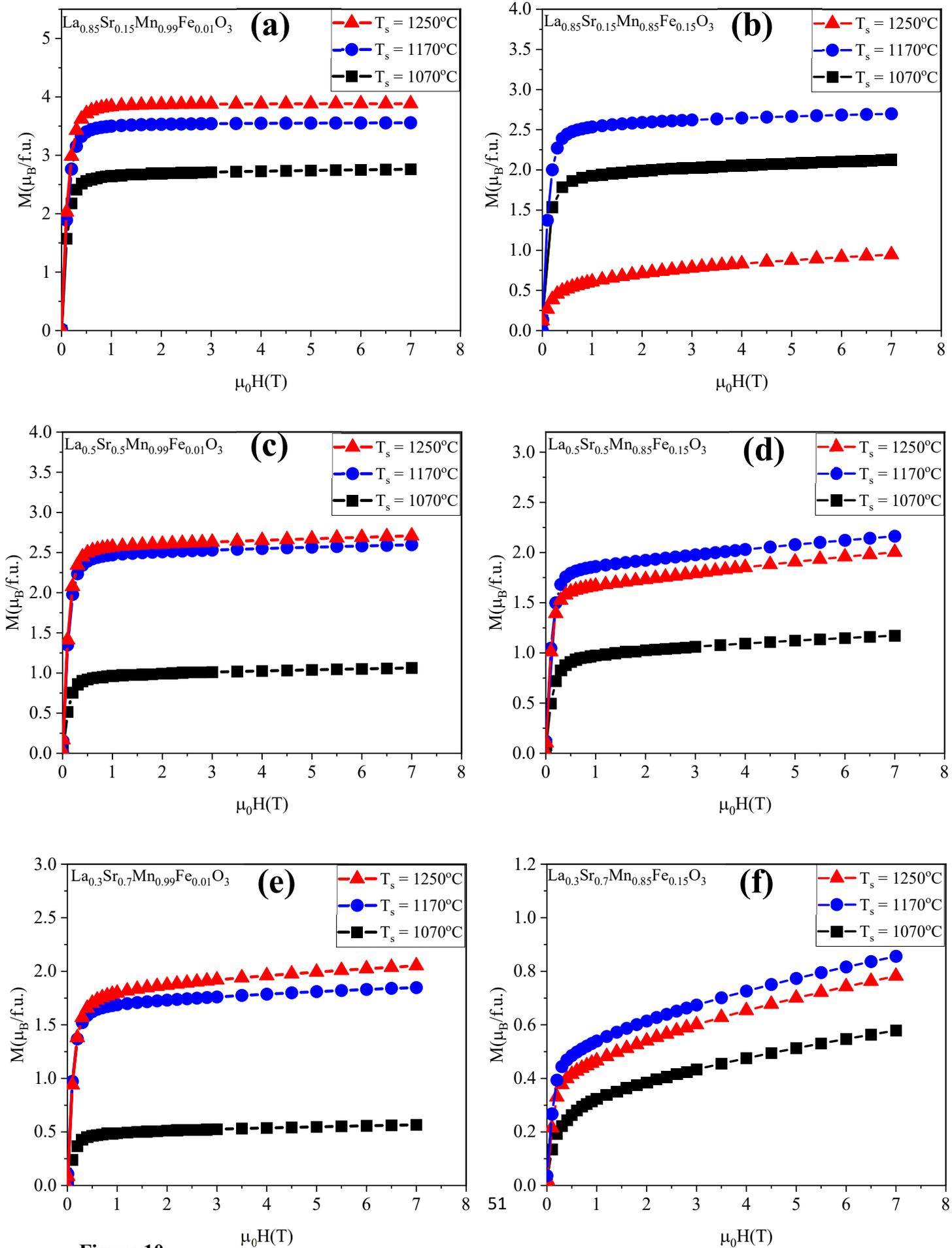

**Figure 10**

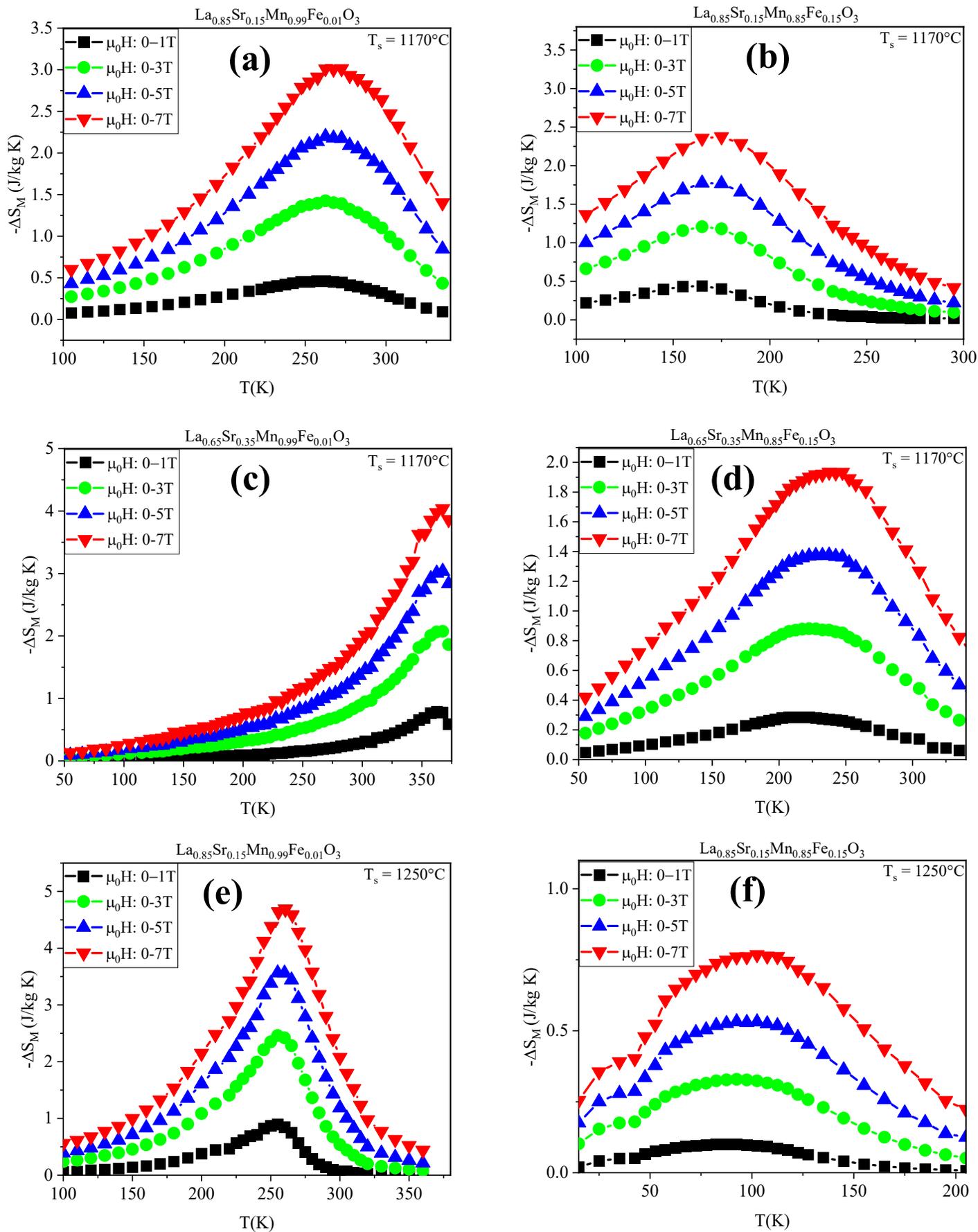

**Figure 11**

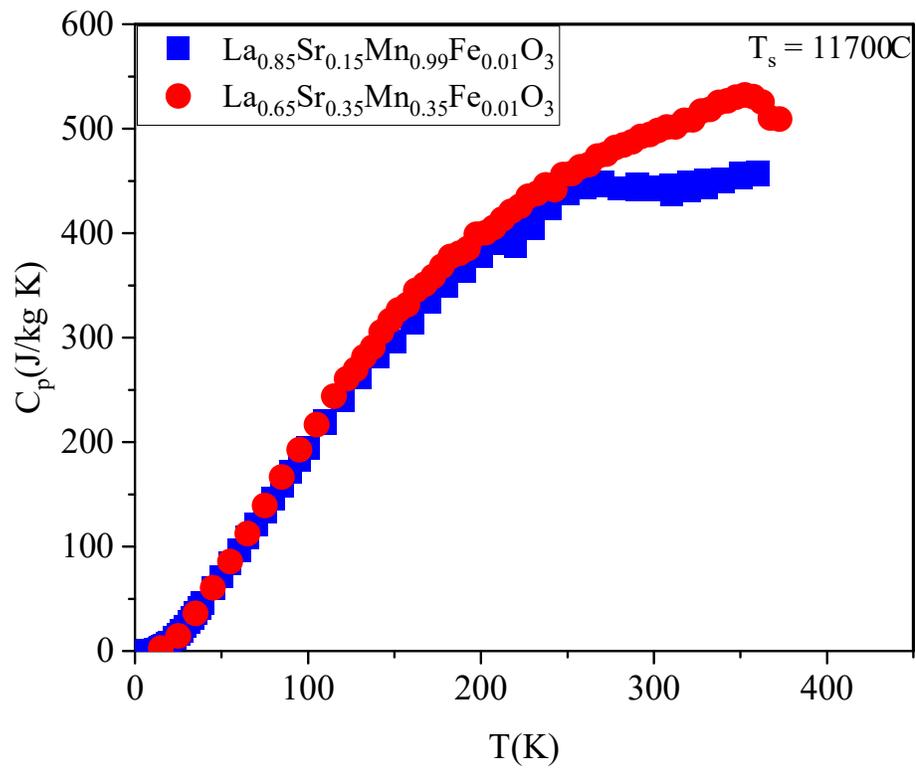

**Figure 12**

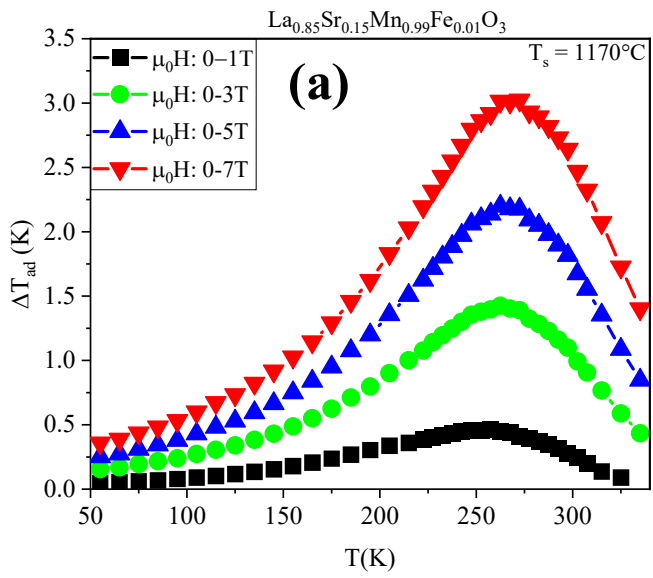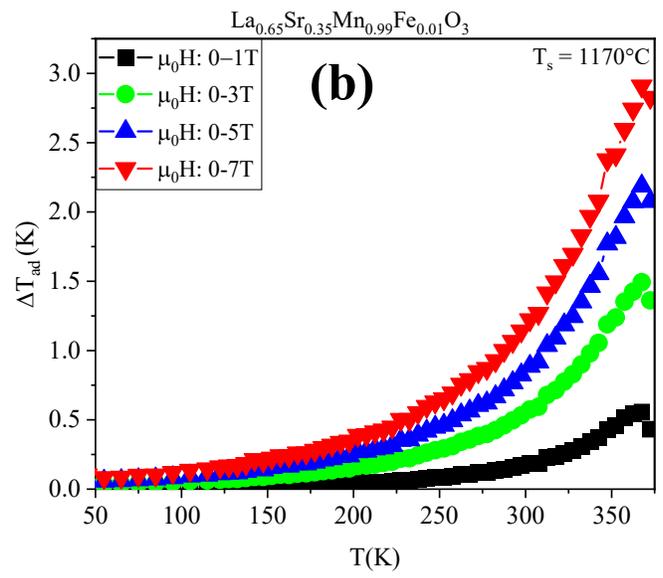

**Figure 13**

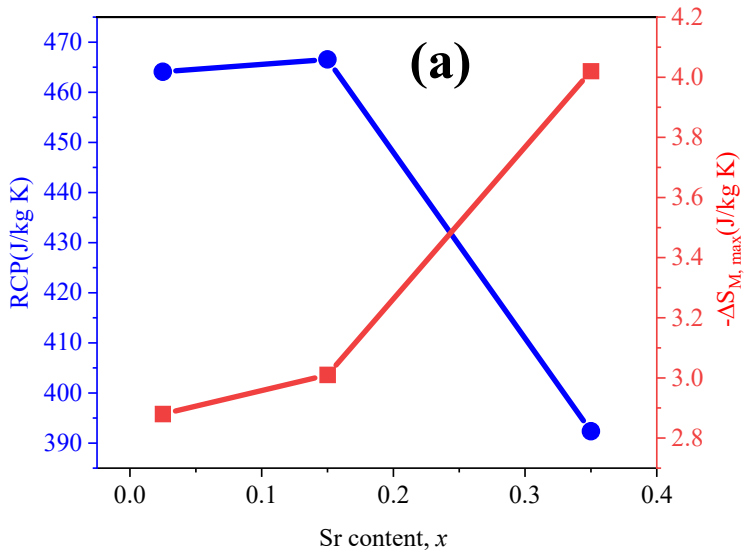 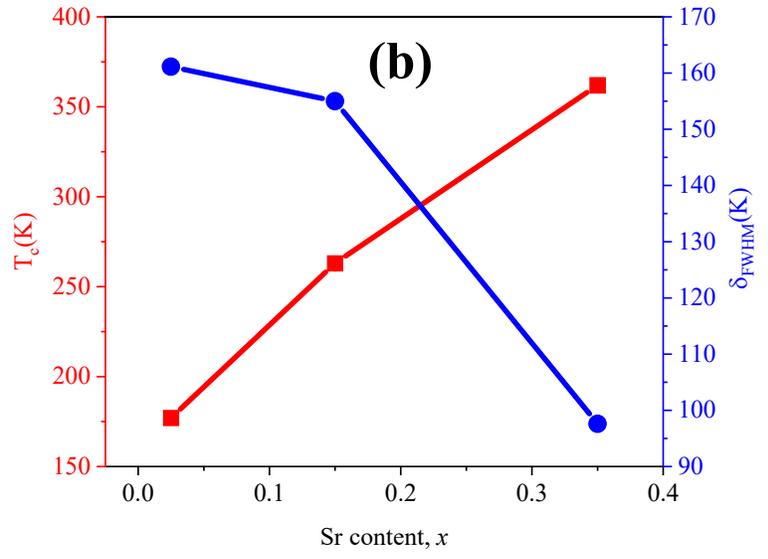

**Figure 14**

# Supplementary materials for: Influence of chemical substitution and sintering temperature on the structural, magnetic and magnetocaloric properties of $La_{1-x}Sr_xMn_{1-y}Fe_yO_3$

N. Brahiti [1], M. Balli [2], M. Abbasi Eskandari [1], A. El Boukili[3,4], P. Fournier [1]

[1] Institut quantique, Regroupement québécois sur les matériaux de pointe et Département de physique, Université de Sherbrooke, Sherbrooke, J1K 2R1, Québec, Canada

[2] AMEEC Team, LERMA, College of Engineering & Architecture, International University of Rabat, Parc Technopolis, Rocade de Rabat-Salé, 11100, Morocco.

[3] LaMCScI Laboratory, B.P. 1014, Faculty of science. Mohammad V University in Rabat, Morocco.

[4] Materials and Nanomaterials Center, MAScIR Foundation, B.P. 10100 Rabat, Morocco.

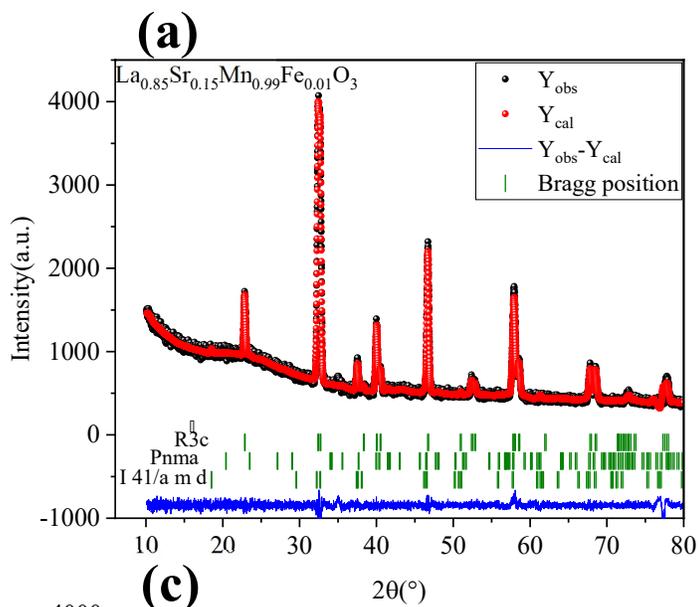
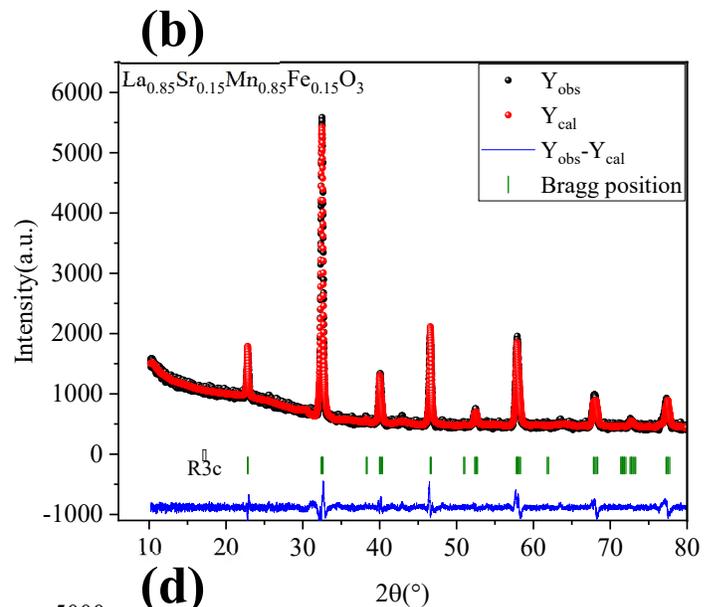
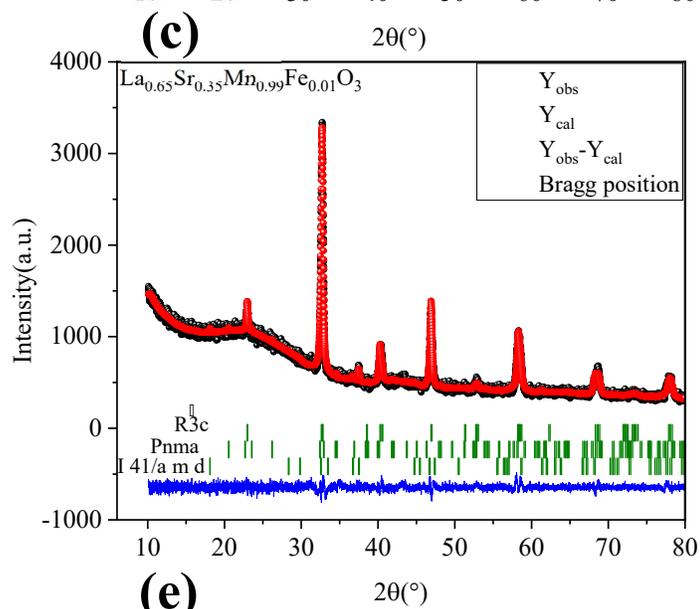
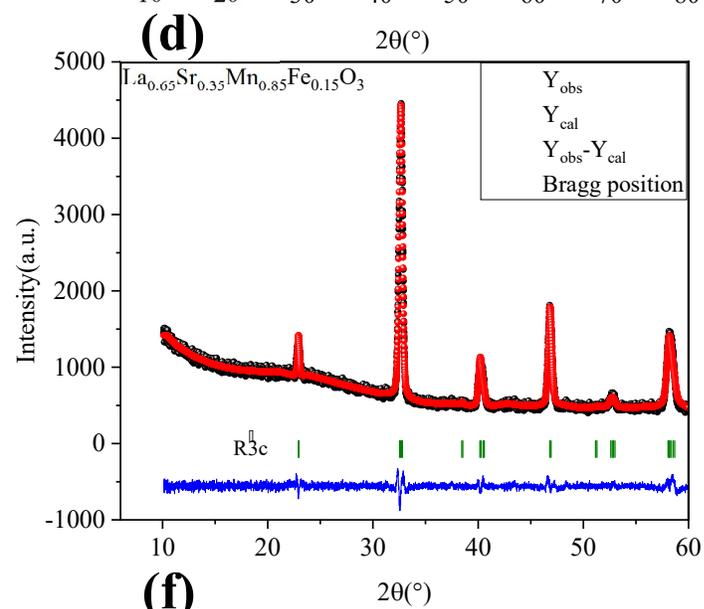
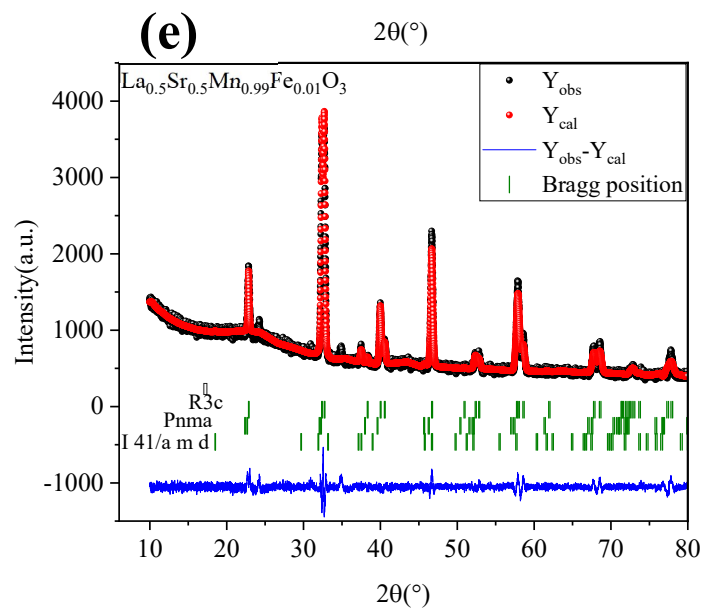
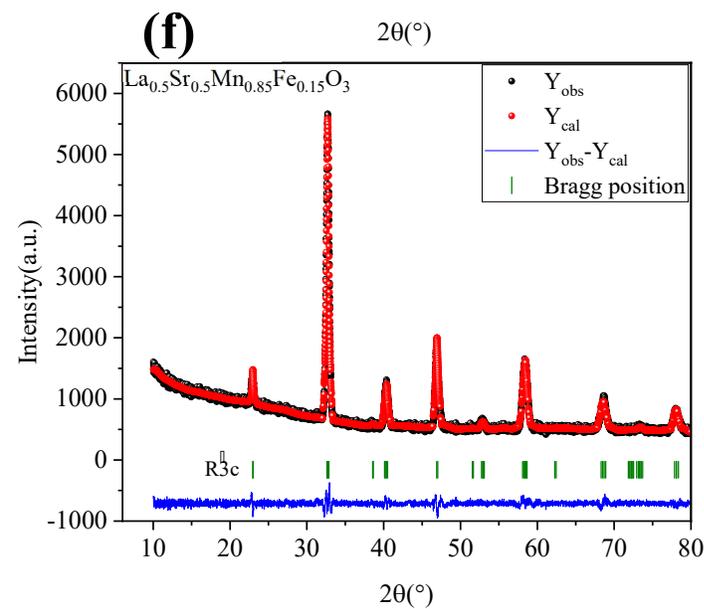

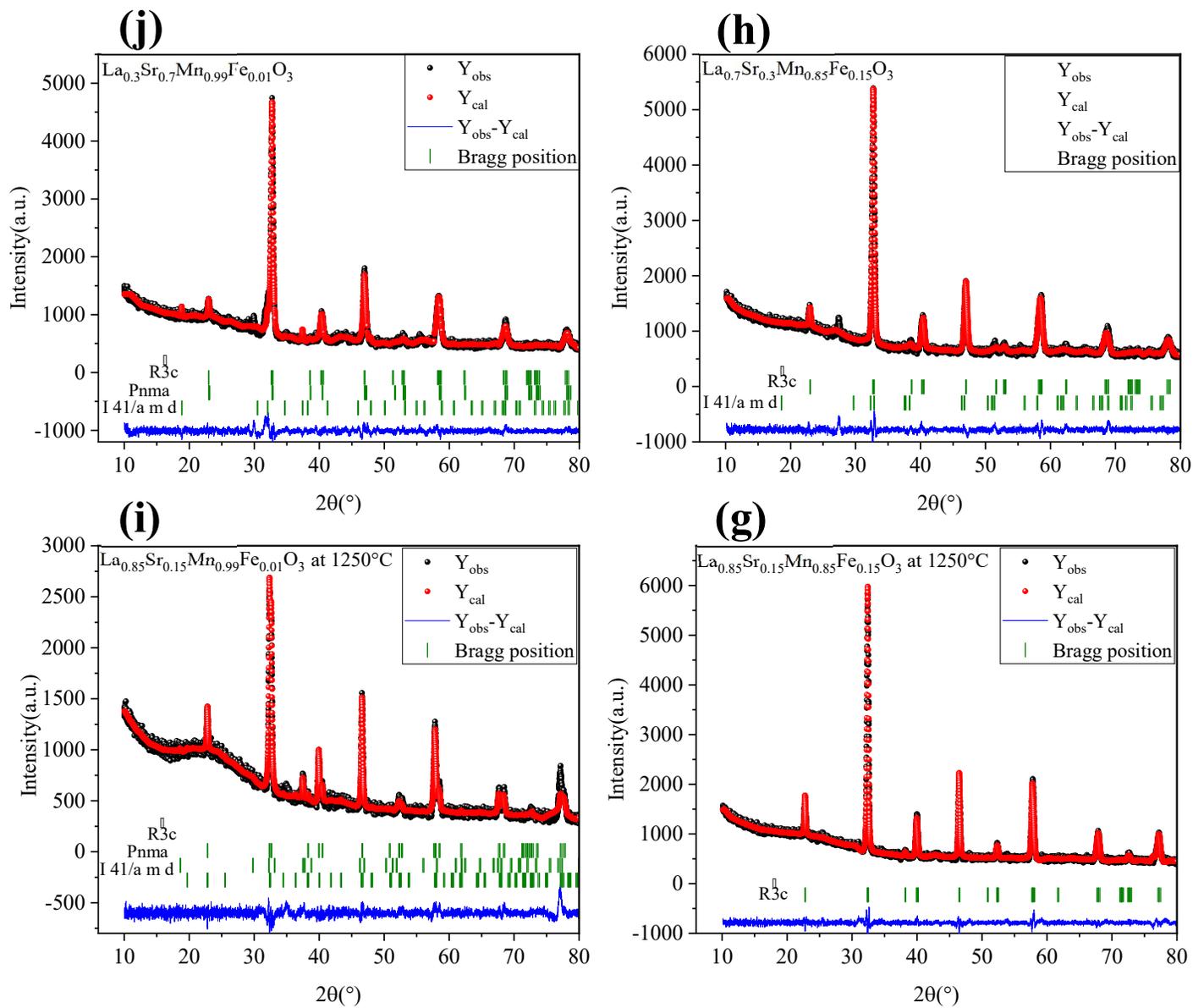

**Figure S1:** Powder XRD patterns along with Rietveld refinements of $La_{1-x}Sr_xMn_{1-y}Fe_yO_3$ ($0.15 \leq x \leq 0.7, y = 0.01, 0.15$) compounds prepared at $T_s = 1170°C$ in (a – f) and prepared at $T_s = 1250°C$ in (i, g)

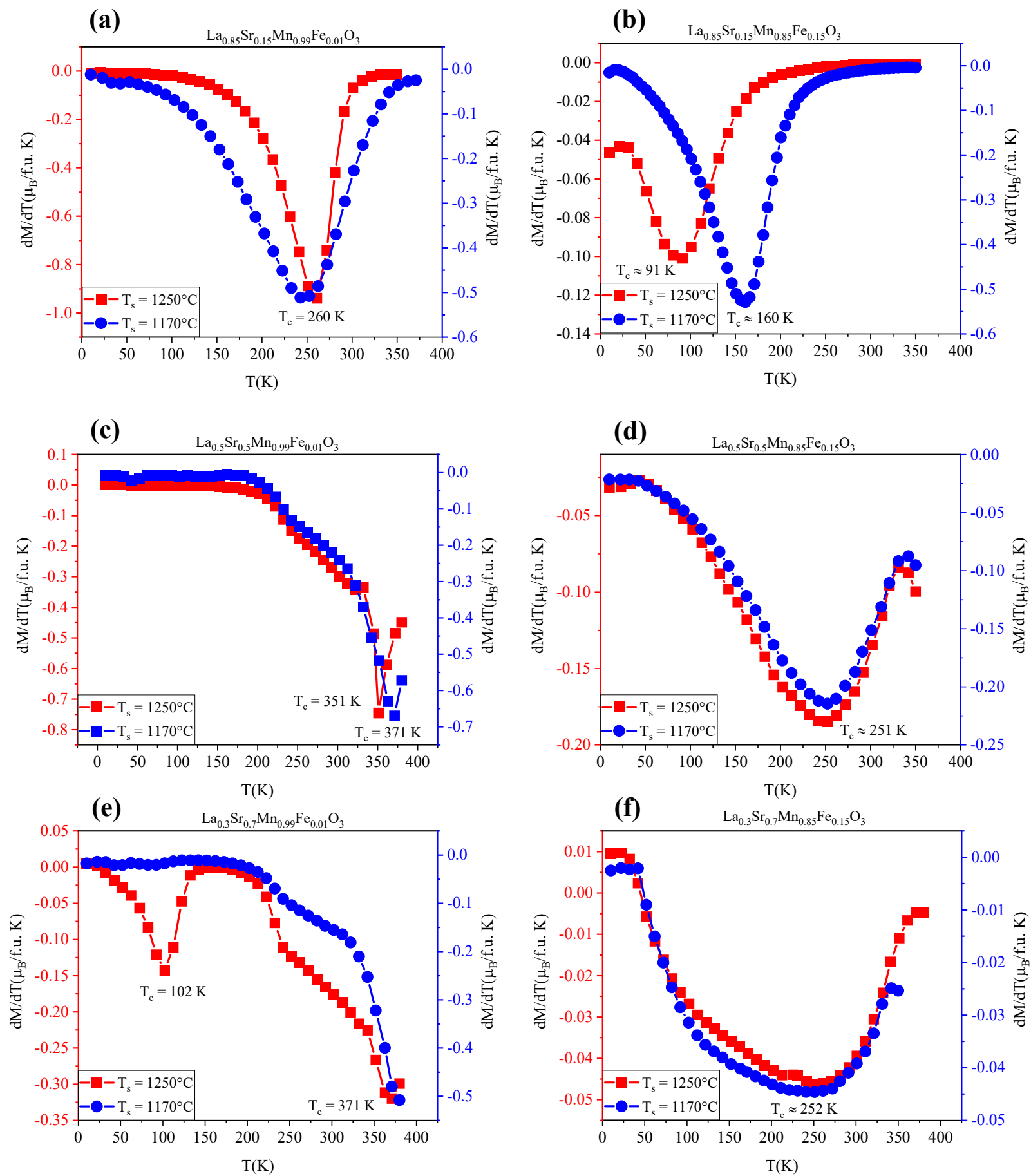

**Figure S2:** The derivative $\frac{dM}{dT}$ as a function of T for $La_{1-x}Sr_xMn_{1-y}Fe_yO_3$ ($x = 0.15, 0.15$ and $0.7$, $y = 0.1$ and $0.15$) at $T_s = 1170°C$ and $T_s = 1250°C$.

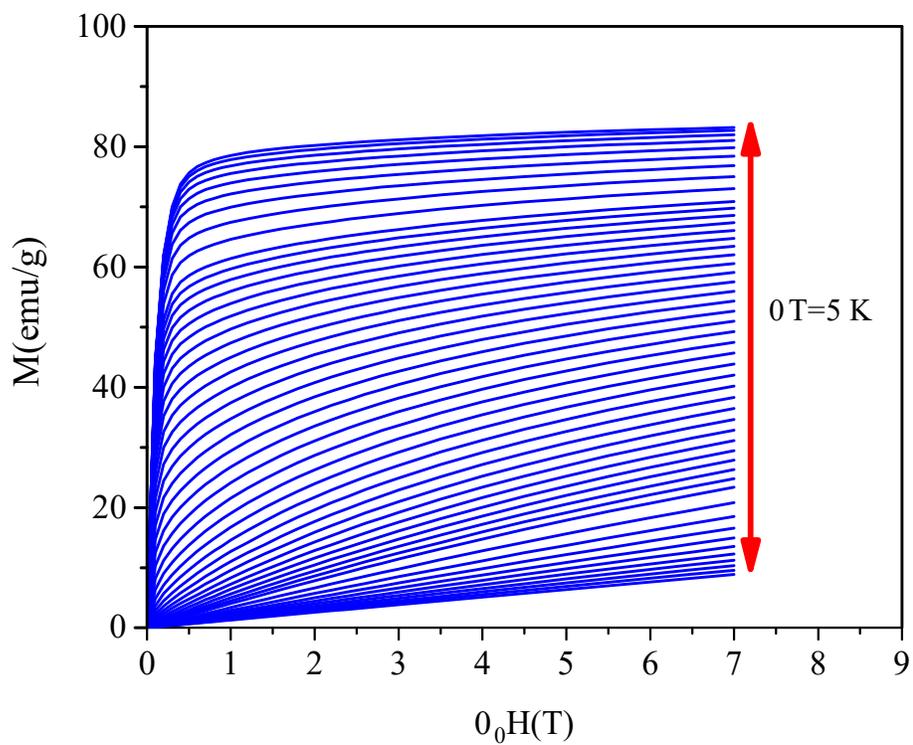

**Figure S3:** Example of isothermal magnetization curves for $La_{0.85}Sr_{0.15}Mn_{0.99}Fe_{0.01}O_3$ sintered at $T_s$ = 1170°C from 5 to 350 K in intervals of 5K used to evaluate the isothermal entropy change.

| Fe content (y) | y = 0.01 | | | | | y = 0.15 | | | | |
|---|---|---|---|---|---|---|---|---|---|---|
| Sr content (x) | 0.025 | 0.15 | 0.35 | 0.5 | 0.7 | 0.025 | 0.15 | 0.35 | 0.5 | 0.7 |
| Space group | R-3c | | | | | R-3c | | | | |
| Biso (Å)² La/Sr Mn/Fe O | 1.107 0.183 0.857 | 1.037 0.862 0.712 | 1.744 0.081 1.464 | 0.052 1.544 0.5 | 0.439 0.473 0.8 | 0.206 0.043 1.026 | 0.694 0.396 0.691 | 0.295 0.386 0.400 | 0.406 0.319 0.412 | 0.331 0.565 0.854 |
| Occupancy La Sr Mn/Fe O | 0.975 0.025 0.978 1.088 | 0.847 0.153 1.006 1.071 | 0.65 0.35 0.986 1.031 | 0.524 0.476 0.940 1.015 | 0.271 0.729 1.048 1.032 | 0.975 0.025 1.004 1.102 | 0.849 0.151 1.005 1.008 | 0.643 0.357 1.003 1.080 | 0.493 0.507 1.018 1.006 | 0.3 0.7 1.001 0.998 |
| Atoms | Coordinates of oxygen ions | | | | | | | | | |
| X (oxygen position) | 0.550 | 0.548 | 0.523 | 0.558 | 0.556 | 0.545 | 0.550 | 0.536 | 0.533 | 0.546 |
| Discrepancy factors | | | | | | | | | | |
| $\chi^2$ | 1.81 | 1.65 | 1.40 | 1.99 | 2.4 | 1.94 | 2.53 | 1.56 | 1.53 | 1.71 |
| $R_p$ | 3.83 | 3.62 | 3.74 | 4.15 | 4.57 | 4.72 | 4.26 | 3.70 | 3.46 | 3.52 |
| $R_{wp}$ | 5.05 | 5.03 | 4.84 | 5.43 | 6.04 | 6.04 | 5.93 | 4.78 | 4.51 | 4.57 |
| $R_{exp}$ | 3.75 | 3.91 | 4.09 | 3.85 | 3.90 | 4.34 | 3.73 | 3.82 | 3.64 | 3.49 |

**Table S1**: Additional parameters extracted from the Rietveld refinements (not presented in Table 1). It includes the isotropic thermal parameters (Biso), the relative oxygen position (X) and the discrepancy factors. All the data are for samples grown at 1170°C.